\documentclass[%
 aip,
 amsmath,amssymb,
 reprint,%
]{revtex4-1}

\usepackage{graphicx}
\usepackage{dcolumn}
\usepackage{bm}

\usepackage[utf8]{inputenc}
\usepackage[T1]{fontenc}
\usepackage{mathptmx}
\usepackage{etoolbox}
\usepackage{float}
\usepackage[usenames,dvipsnames]{xcolor}
\usepackage{cancel}

\newcommand{\bg}{{\bf g}}
\newcommand{\bq}{{\bf q}}
\newcommand{\br}{{\bf r}}
\newcommand{\bk}{{\bf k}}
\newcommand{\al}{\alpha}
\newcommand{\be}{\beta}

\newcommand{\gps}[1]{\textcolor{black}{#1}}
\newcommand{\gpsch}[1]{\textcolor{black}{#1}}
\newcommand{\gk}[1]{\textcolor{black}{#1}}
\newcommand{\ggr}[1]{\textcolor{black}{#1}}
\newcommand{\gr}[1]{\textcolor{black}{#1}}
\newcommand{\gb}[1]{\textcolor{black}{#1}}
\makeatletter
\def\@email#1#2{%
 \endgroup
 \patchcmd{\titleblock@produce}
  {\frontmatter@RRAPformat}
  {\frontmatter@RRAPformat{\produce@RRAP{*#1\href{mailto:#2}{#2}}}\frontmatter@RRAPformat}
  {}{}
}%
\makeatother

\begin{document}

\preprint{AIP/123-QED}

\title{Elasticity in crystals with high density of local defects : insights from ultra-soft colloids}
\author{Saswati Ganguly}
 \affiliation{Soft Condensed Matter Theory, Fachbereich Physik, Universit\"at Konstanz, 78457 Konstanz, Germany}
\author{Gaurav Prakash Shrivastav}%
\affiliation{ 
Institut f\"ur Theoretische Physik, TU Wien, Wiedner Hauptstrasse 8-10, A-1040 Wien, Austria
}%
\author{Shang-Chun Lin }
\affiliation{%
Institut f\"ur Angewandte Physik, Eberhard Karls Universit\"at T\"ubingen, Auf der Morgenstelle 10, 72076 T\"ubingen, Germany
}%
\author{Johannes H\"aring}
 \affiliation{Soft Condensed Matter Theory, Fachbereich Physik, Universit\"at Konstanz, 78457 Konstanz, Germany}
 \author{Rudolf Haussmann}
 \affiliation{Soft Condensed Matter Theory, Fachbereich Physik, Universit\"at Konstanz, 78457 Konstanz, Germany}
\author{Gerhard Kahl}%
\affiliation{ 
Institut f\"ur Theoretische Physik, TU Wien, Wiedner Hauptstrasse 8-10, A-1040 Wien, Austria
}%
\author{Martin Oettel}
\affiliation{%
Institut f\"ur Angewandte Physik, Eberhard Karls Universit\"at T\"ubingen, Auf der Morgenstelle 10, 72076 T\"ubingen, Germany
}%
\author{Matthias Fuchs}
 \affiliation{Soft Condensed Matter Theory, Fachbereich Physik, Universit\"at Konstanz, 78457 Konstanz, Germany}

\date{\today}

\begin{abstract}
In complex crystals close to melting or at finite temperatures, different types of defects are ubiquitous and their role becomes relevant in the mechanical response of these solids. Conventional elasticity theory fails to provide a microscopic basis to include and account for the motion of point-defects in an otherwise ordered crystalline structure. We study the elastic properties of a point-defect rich crystal within a first principles theoretical framework derived from microscopic equations of motion. This framework allows us to make specific predictions pertaining to the mechanical properties which we can validate through deformation experiments performed in Molecular Dynamics simulations.
\end{abstract}

\maketitle

\section{\label{intro}Introduction}
Solids are characterised by their resistance to external deformations. From a symmetry perspective, rigidity in crystalline solids is attributed to spontaneous breaking of continuous translational invariance which transcribes to particle arrangements with long range order~\cite{Ashcroft_Mermin1976}. The thermodynamics of the reversible linear elastic response stated in Hooke's law is understood by focusing on macroscopic length-scales and describing the deformations as small deviations away from an equilibrium un-deformed crystalline structure~\cite{LL_ET}. Now, when an equilibrium system with a large number of microscopic degrees of freedom is subjected to a small perturbation, most of the degrees relax quickly back to the equilibrium and their relaxation time scales depend on the details of the microscopic interactions in the system. The bulk thermodynamics of the system, however, is governed by a few collective modes whose relaxation times are proportional to some power of their wavelength implying diverging relaxation time-scales for perturbations of system spanning length-scales. These slow hydrodynamic modes can be propagative like the two sound propagation modes in a single component fluid where the frequency vanishes linearly with the wave number of this mode. It can also be diffusive, where the frequency of the mode vanishes quadratically with its wave number, like in case of heat transport and the two transverse shear waves in a fluid. In case of homogeneous systems like a simple fluid, the slow hydrodynamic variables follow from the conservation laws in the system. The crystalline solid, on the other hand, has three additional slow modes called the Nambu-Goldstone modes arising because of the spontaneous breaking of the three continuous spatial translations. Phenomenological hydrodynamic theories~\cite{Martin_Parodi_Pershan,Fleming_Cohen} in the seventies established these eight slow collective modes as the basis for understanding the macroscopic mechanical response in crystalline solids. One of these hydrodynamic modes is assigned to the diffusion of point defects. These theories, unfortunately, do not provide any further insight regarding the microscopic interactions at atomic length-scales. 

Familiar continuum elasticity theory~\cite{chaikin_lubensky_1995,LL_ET, Doghri1964} has the strain tensor as the central observable quantity that associates the bulk deformation with the particle displacements. In this case the displacement fields ${\bf u}_{i}(t)={\bf r}_{i}(t)-{\bf R}_{i}$, for the particles at ${\bf r}_{i}(t)$ at a given time $t$ in a crystal at finite temperature, are defined with respect to reference lattice positions ${\bf R}_{i}$. The reference ${\bf R}_{i}=\langle {\bf r}_{i}(t)\rangle$ represents the ideal lattice structure with its perfect periodicity manifesting the long-range translational order expected in a crystalline solid. The one-to-one mapping of particles to lattice positions also trivially connects the change of a coarse grained density field to the divergence of the displacement field $\delta  n(\br,t)=-n_{0}\nabla \cdot {\bf u}(\br,t)$. 
Here, $n_{0}=N/V$ is the average number density for a system of $N$ particles and $V$ volume, and the coarse graining is over many unit cells in the continuum picture. This description, however, does not account for the diffusion of point-defects. This omission is problematic because phenomenological understanding~\cite{Martin_Parodi_Pershan,Fleming_Cohen} of long-wavelength and low frequency excitations, in ordered solids predict the defect diffusion to be one of the eight hydrodynamic modes. Therefore, in a real crystal with point-defects, lattice deformation is not the sole contributor to density changes. Fluctuations in the concentration of point-defects $\delta c_{d}(\br,t)$ (likewise coarse grained) also leads to fluctuations in densities. 
\begin{equation}\label{ansatz1_real}
    \delta n(\br,t)=-n_{0}\nabla \cdot {\bf u}(\br,t)-\delta c_{d} (\br,t)
\end{equation}
The defect density $c_{d}(\br,t)$ is positive for vacancies and negative for interstitials. Starting with these interpretations of density, displacement fields, defect density, and the Hamiltonian governing the microscopic atomic motions, Ref.~\onlinecite{Walz2010} derived a theory for the linear elastic response of local-defect rich non-ideal crystals. The connection between this non-equilibrium statistical mechanics perspective and the thermodynamics of deformation was further developed in Ref.~\onlinecite{Haring2015}.
In particular, in these works the coarse grained fields for displacement and defect density were connected with changes in a \textit{microscopic} density $\delta\rho(\br,t)$ and the response coefficients for static elasticity and elastic waves could be expressed in terms of the microscopic density and microscopic two--point functions \gr{(density--density and direct correlation function)} for the solid in the undeformed equilibrium state. In a recent paper, the formalism was applied to the hard sphere system\cite{lin2021} which is \textit{defect--poor}. In this system, the effect of defects is manifest in the crystalline direct correlation function but the effect on the conventional elastic constants is minor due to cancellations in the expression for the elastic response.

In this contribution, we examine how the explicit implementation of this framework for a model \textit{point-defect rich} crystalline solid leads to insights regarding the role of local-defects in determining the mechanical response. Our analysis involved in making theoretical predictions for different properties associated with elasticity in crystals, also brings out some important assumptions. The implications of these assumptions are further explored when we compare our predictions with atomistic Molecular Dynamics simulations.

The rest of the paper is organised as follows:
Section~\ref{theory} is devoted to developing and explaining the theoretical framework. Section~\ref{Simulation} describes the model system (Section~\ref{sec_model}) we chose to implement our theory in and the Molecular Dynamics simulations (Section~\ref{subsec:sim_cluster}) performed to validate the theoretical predictions. Most of Section~\ref{theory} is not specific for a particular ordered system or crystal structure but instead provides the general technical details of the theory. Section~\ref{theory} contains three subsections each dealing with one specific facet of the theoretical framework. For the sake of completeness, Section~\ref{MZ_theo} recapitulates the microscopic basis for understanding the reversible linear elastic response in point-defect rich crystals as proposed in Ref.~\onlinecite{Walz2010}. One of the key outcomes of this section is the wave vector ($\bq$) dependent dynamical matrix $\Lambda_{\al\be}(\bq)$ which appears in the wave equations for the displacement and the momentum density fields. Section~\ref{thermodynamics} shows how the quantities emerging from the microscopic perspective relates to standard concepts of linear elasticity theory. Here we discuss the different definitions and the corresponding symmetry considerations of the coefficients of elasticity. In order to make explicit predictions regarding mechanical properties of any specific system, the equations summarised in Section~\ref{MZ_theo} require input in the form of static correlation functions, in particular the direct correlation function, which are conveniently obtainable from classical density functional theory. Section~\ref{Ec_cdft} derives generalized elastic coeffcients and summarizes final expressions for the various constants of elasticity. 
Subsequently, Section~\ref{sec_model} introduces the definition and density functional for our model system of a point defect--rich cluster crystal. 
The implementation of the Molecular Dynamics simulations performed to emulate deformation experiments of this system is described in Section~\ref{subsec:sim_cluster}. Section~\ref{results} discusses the quantitative evaluations of the linear elastic properties. 
Section~\ref{Ec_theo} estimates various constants of elasticity using the ideas developed in Section~\ref{theory}. Here the assumptions involved in these analysis are commented upon and contrasted with available examples in the literature. The theoretical predictions and insights, so obtained, are validated and compared to results obtained from Molecular Dynamics simulations in Section~\ref{sec_compare}.


\section{\label{theory}The theory}
\subsection{\label{MZ_theo}The microscopic basis}

With the aim of describing the thermodynamics of deformation for the crystalline state with a few slowly relaxing relevant variables, the Mori-Zwanzig projection operator formalism~\cite{Zwanzig2001,Forster} has been used. This concept establishes the connection between the macroscopic mechanical response of the crystal to the microscopic Hamilton equations. Deriving the equations of motions of the fields, that recover the continuum description in the hydrodynamic limit, requires the identification of the relevant variables.

As the crystalline solid is characterised by its spontaneously broken continuous translational symmetry, one needs to consider additional hydrodynamic variables besides the conserved quantities. The variables with long-ranged equilibrium correlations arising due to the spontaneous breaking of symmetry are identified relying on the Bogoliubov inequality~\cite{Forster}. For crystals, this variant of Schwarz’s inequality involving the momentum and the mass density fluctuations, was proposed by Wagner~\cite{Wagner}. Thus in the case of an isothermal crystal at temperature $T$ and density $n_{0}$, the fields associated with the mass and momentum conservation and the restoration of the broken symmetry, constitute the set of hydrodynamic variables. 

\gr{We define the microscopic particle density operator by $\hat{\rho}(\br,t)=\sum_{i=1}^{N}\delta\left(\br-\br_{i}(t)\right)$. Its fluctuation in 
reciprocal space is given by}
\begin{eqnarray}
&\delta \hat{\rho}(\bg+\bq,t)=\hat{\rho}(\bg+\bq,t)-n_{\bg}V\delta_{\bq 0} .
\end{eqnarray}
Here the spatial Fourier transform of the particle density \gr{operator} is given by
\begin{equation}
    \hat{\rho}(\bk,t)=\sum_{i=1}^{N}e^{-i{\bf k} \cdot \br_{i}(t)}=\sum_{i=1}^{N}e^{-i(\bg+\bq) \cdot\br_{i}(t)} .
\end{equation}
\gr{A microscopic field for the density fluctuations is defined by averaging $\delta\hat\rho(\bk,t)$ over a
suitable (time--dependent) many--body distribution,
\begin{equation}
\delta \rho(\bk,t)=\langle \delta \hat{\rho}(\bk,t)\rangle^\text{td}.    
\end{equation}
The wave vector $\bk=\bg+\bq$ is split into the reciprocal lattice vector $\bg$ and the wave vector lying within the first Brillouin zone, $\bq$. 
For a crystal in equilibrium, averages of $\hat\rho(\bg+\bq,t)$ have only contributions at reciprocal lattice vectors, defining the}  
Bragg peak amplitudes $n_{\bg}$,
\begin{equation}
    n_{\bg}=\dfrac{1}{V}\langle\hat{\rho}(\bg,t)\rangle=\dfrac{1}{V}\left\langle \sum_{i=1}^{N}e^{-i\bg \cdot \br_{i}(t)}\right\rangle
\end{equation}
representing the spontaneous breaking of translational invariance in ordered systems. \gr{These} also serve as the hydrodynamic variable associated with conservation of mass. The choice of $n_{\bg}$ as a slow, \gr{relevant density variable (therefore denoted by ``$n$'' as opposed of ``$\rho$'' for a 
general microscopic density)} is justified using the Bogoliubov inequality~\cite{Forster} which provides an argument that the inverse density correlations vanish as $\propto k^{2}$ for wave vectors close to all non-zero reciprocal lattice vectors $\bg\neq0$. 

\gr{
Associated slow, relevant fluctuations in these $n_{\bg}$ will be averages over 
$\delta\hat\rho(\bg+\bq,t)$ in the linear response regime\cite{Walz2010} and for small $\bq$, denoted by
\begin{equation}
\delta n_{\bg}(\bq,t)=\langle \delta \hat{\rho}(\bg+\bq,t)\rangle^\text{lr},    
\end{equation}
for which equations of motion in the Mori--Zwanzig formalism can be derived (see below).
}


The other set of slow variables \gr{derive from the operator for} the momentum density components $\hat{j}_{\al}(\br,t)=\sum_{i=1}^{N}p_{\alpha}(\br_{i})\delta\left(\br -\br_{i}(t)\right)$
associated with the conservation of linear momentum
\begin{eqnarray}
\partial_{t}\hat{j}_{\al}(\bk)-ik_{\beta}\hat{\sigma}_{\alpha\beta}(\bk)=0
\end{eqnarray}
where $\hat{\sigma}_{\alpha\beta}$ is the stress tensor.
The Einstein convention of summation over repeated indices has been used here and in all subsequent equations.
\gr{The associated slow, relevant fluctuations are given by 
\begin{equation}
    \delta j_{\al}(\bq,t)=\langle \hat{j}_{\alpha}(\bq,t)\rangle^\text{lr}.
\end{equation}
}

Following the Mori-Zwanzig formalism, one arrives at the linear equations of motion for such a set of selected relevant variables $\{\hat{A}_{i}(t)\}$. Dropping the memory kernels, the dissipation-less reversible equations of motion for small changes in
\gr{the time-dependent} averages$\{\hat{A}_{i}(t)\}$ away from their equilibrium values are obtained in terms of correlation functions evaluated in the unperturbed equilibrated system~\cite{Zwanzig2001,Forster}
\begin{align}
    \partial_{t}\langle \delta \hat{A}_{i}(t)\rangle^\text{\gr{lr}}&=i\Omega^{*}_{ik}\langle\delta \hat{A}_{k}(t)\rangle^\text{\gr{lr}}\nonumber\\
    &=i\left[\langle\delta \hat{A}^{*}_{i}\mathcal{L}\delta \hat{A}_{j}\rangle\langle\delta \hat{A}^{*}_{j}\delta \hat{A}^{*}_{k}\rangle^{-1}\right]^{*}\langle\delta \hat{A}_{k}(t)\rangle^\text{\gr{lr}} .
\end{align}
The Liouville operator $\mathcal{L}$ (entering the definition of the frequency matrix $\Omega_{ik}$) and the averages represented by the angular brackets correspond to the canonical ensemble. Now for the chosen set of relevant variables \gr{$\delta n_{\bg}(\bq,t)$ and $\delta j_{\al}(\bq,t)$},
this route leads to the dissipation-less isothermal equations of motion~\cite{Walz2010}
\gr{
\begin{subequations}\label{eq_EOM}
\begin{align}
\partial_{t}\delta n_{\bf g}({\bf q},t)
&=i\left(\dfrac{\langle\delta \hat{\rho}^{*}({\bf g+q})\mathcal{L} \hat{j}_{\alpha}({\bf q})\rangle}{\langle \hat{j}^{*}_{\alpha}({\bf q}) \hat{j}_{\beta}({\bf q})\rangle}\right)^{*}\delta j_{\beta}({\bf q},t)\nonumber\\
&=-i\dfrac{n_{\bg}}{mn_{0}}(g+q)_{\alpha}\delta j_{\alpha}({\bf q},t)\\
\partial_{t}\delta j_{\alpha}({\bf q},t)&=i\sum_{\bg',\bg}\left(\dfrac{\langle  \hat{j}^{*}_{\alpha}({\bf q})\mathcal{L}\delta \hat{\rho}({\bf g'+q}) \rangle}{\langle\delta \hat{\rho}^{*}({\bf g'+q})\delta \hat{\rho}({\bf g+q}) \rangle}\right)^{*}\delta n_{\bf g}({\bf q},t)\nonumber\\
&=-i\sum_{{\bf g}',{\bf g}}(g'+q)_{\al}n^{*}_{\bf g'}J^{*}_{{\bf g}'{\bf g}}(\bq)\delta n_{\bf g}({\bf q},t)
\end{align}
\end{subequations}
{\tt no $\delta$ before $\hat j$.}
}

The inverse density correlation matrix $J_{\bg'\bg''}$ appearing in the equation of motion for the momentum density (Eq.~\ref{eq_EOM}b) is defined as follows
\begin{align}
 \label{eq_J}
Vk_{B}T\delta _{\bg\bg''}=\sum_{\bg'}\langle \delta \hat{\rho}^{*}_{\bg}(\bq,t)\delta \hat{\rho}_{\bg'}(\bq,t)\rangle J_{\bg'\bg''}(\bq) 
\end{align}
where $k_{B}$ is the Boltzmann constant. \gr{Note that the averages are defined in equilibrium such that the density fluctuation correlator and the matrix $J$ are purely static objects.}

The time derivative of Eq.~\ref{eq_EOM}b followed by substitution using Eq.~\ref{eq_EOM}a leads to the wave equation 
\begin{align}\label{eq_wave1}
    \partial^{2}_{t}\delta j_{\alpha}({\bf q},t)&=-\dfrac{1}{mn_{0}}\Lambda_{\alpha\beta}({\bf q})\delta j_{\beta}({\bf q},t)\nonumber\\
    &=-\dfrac{1}{mn_{0}}\sum_{{\bf g}',{\bf g}}(g'+q)_{\al}n^{*}_{\bf g'}J^{*}_{{\bf g}'{\bf g}} n_{\bf g}(g+q)_{\beta}\delta j_{\beta}({\bf q},t)
\end{align}
and the introduction of the dynamical matrix $\Lambda_{\alpha\beta}(\bq)$.

\begin{align} \label{eq_Lambda}
  \Lambda(\bq)&=\sum_{\bg,\bg'}(g'+q)_{\alpha}n^{*}_{\bg'}J^{*}_{\bg'\bg}n_{\bg}(g+q)_{\beta}
\end{align}
This equation contains all the information pertaining to elastic coefficients and hence also the  sound velocities in our crystalline solid of interest~\cite{Ashcroft_Mermin1976}.

In order to interpret the above equations, the following ansatz~\cite{Walz2010} is required to establish a connection between the microscopic fluctuations $\delta n_{\bg}(\bq,t)$ and the coarse-grained elastic field of displacement fluctuations $\delta u_{\alpha}(\bq,t)$
\begin{equation}\label{ansatz1}
    \delta n_{\bg}(\bq,t)=-in_{\bg}(g+q)_{\al}\delta u_{\al}(\bq,t)+\dfrac{n_{\bg}}{n_{0}}\delta c_{d}(\bq,t) .
\end{equation}
Relation \ref{ansatz1} can be rationalised by realising that for a perfect crystal, the density fluctuations originate solely from the divergence of the displacement field.
Here the point-defects are accounted for by introducing the term proportional to $\delta c_{d}(\bq,t)$. 
Nevertheless, \gr{already for time--independent deformations} this is an approximation missing contributions from the full pair correlation function~\cite{future_RH}.
Inserting Eq.~\ref{ansatz1} into Eq.~\ref{eq_EOM}b and using the definition of the velocity field as the time derivative of the displacement field ${\bf u}$,  the wave equation for the displacement field is obtained. 

\begin{equation}\label{wave_u}
    \partial^{2}_{t}\delta u_{\al}(\bq,t)=-\dfrac{1}{mn_{0}}\Lambda_{\al\be}(\bq)\delta u_{\be}(\bq,t)-\dfrac{1}{mn_{0}}V_{\al}(\bq)\delta c_{d}(\bq,t) .
\end{equation}
Here the term $V_{\al}(\bq)$ is given by the following relation. 
\begin{align}
    V_{\al} ({\bf q})=-\dfrac{i}{n_{0}}\sum_{\bg'\bg}(g'+q)_{\al}n^{*}_{\bg'}J^{*}_{\bg'\bg}(\bq)n_{\bg}. 
\end{align}
$\Lambda_{\al\be}(\bq)$ is the dynamical matrix at a constant concentration of point defects. The following Section~\ref{thermodynamics} summarises the connection between $\Lambda_{\al\be}(\bq)$ and the thermodynamics associated with the mechanical properties of solids.

\subsection{\label{thermodynamics}The connection to thermodynamics}
\subsubsection{\label{th_fen}The free energy and the constants of elasticity}

The linear, isothermal, elastic constants of materials can be defined as the second order strain derivatives of the Helmholtz free energy. The Helmholtz free energy, $F$, of an equilibrium crystal at temperature $T$ with volume $V$ and pressure \gr{$P=-\partial F/\partial V$} can be expanded~\cite{WALLACE1970} in terms of symmetrised Lagrangian strains $\eta_{\al\be}=\dfrac{1}{2}\left(\nabla_{\al}u_{\be}+\nabla_{\be}u_{\al}+\nabla_{\al}u_{\gamma}\nabla_{\be}u_{\gamma}\right)$, measured with respect to the equilibrium lattice
\begin{equation}\label{F_eta}
    F(\eta_{\al\be})=F(0)+V\tau_{\al\be}\eta_{\al\be}+\dfrac{1}{2}VC_{\al\be\gamma\delta}\eta_{a\be}\eta_{\gamma\delta}+\dots .
\end{equation}

 In the absence of an external, deforming field, the stress tensor $\tau_{\al\be}$ measured at the equilibrium reference lattice, is a diagonal matrix with $(-P)$ as the diagonal terms. The isothermal elastic constant tensor $C_{\al\be\gamma\delta}$, obtained as the second derivative of the free energy 
\begin{equation}
    C_{\al\be\gamma\delta}=V^{-1}\left(\dfrac{\partial^{2}F}{\partial \eta_{\al\be}\partial\eta_{\gamma\delta}}\right)_{T,\eta'\ne \eta}
\end{equation}
has the full Voigt symmetry allowing the following representation of paired indices $\al=1,2,3,4,5,6 \text{ for } \al\be=11,22,33,23 \text{ or } 32,13 \text{ or } 31,12 \text{ or } 21$ respectively in three dimensions.  Alternatively, the free energy can be written in terms of the un-symmetrised displacement gradients $u_{\al\be}=\nabla_{\be}u_{\al}$ or the symmetrised linear strain $\epsilon_{\al\be}=\dfrac{1}{2}(u_{\al\be}+u_{\al\be})$ by substituting the symmetrised Lagrangian strain $\eta_{\al\be}$ in Eq.~\ref{F_eta}
\begin{subequations}
\begin{align}
    \eta_{\al\be}&=\dfrac{1}{2}\left(u_{\al\be}+u_{\be\al}+u_{\gamma\al}u_{\gamma\be}\right)\\
    &=\epsilon_{\al\be}+\dfrac{1}{2}\left(\epsilon_{\gamma\al}\epsilon_{\gamma\be}+\omega_{\gamma\al}\omega_{\gamma\be}
    \right) .
\end{align}
\end{subequations}
For all our comparisons with simulated systems, we choose symmetric deforming fields with no torque. Therefore, for the subsequent analysis we ignore the anti-symmetric part of the strain tensor $\omega_{\al\be}=\dfrac{1}{2}(u_{\al\be}-u_{\be\al})$. The free energy expansions in terms of $\epsilon_{\al\be}$ 
or  $u_{\al\be}$
introduces two 
alternative elasticity tensors
$B_{\al\be\gamma\delta}$ and $A_{\al\be\gamma\delta}$ 
with second--order contributions to the free energy in the form of
$(V/2)B_{\al\be\gamma\delta}\epsilon_{\al\be}\epsilon_{\gamma\delta}$ and
$(V/2)A_{\al\be\gamma\delta}u_{\al\be}u_{\gamma\delta}$, respectively.
In case of an equilibrium crystal at isotropic pressure $P$, they are related to the $C_{\al\be\gamma\delta}$ as follows
\begin{subequations}\label{A_B_C_rel}
\begin{align}
&A_{\alpha\beta\gamma\delta}=-P\delta_{\beta\delta}\delta_{\alpha\gamma}+C_{\alpha\beta\gamma\delta}\\
&B_{\alpha\beta\gamma\delta}=
-P\left[\delta_{\beta\delta}\delta_{\alpha\gamma}
+\delta_{\alpha\delta}\delta_{\beta\gamma}-\delta_{\alpha\beta}\delta_{\gamma\delta}\right]+C_{\alpha\beta\gamma\delta} .
\end{align}
\end{subequations}

On the one hand, the tensor $A_{\al\be\gamma\delta}$ appears in the wave equations for the displacement fields. Wave propagation experiments measure the wave velocities or the eigenvalues of the matrix
\begin{equation}
\Lambda_{\al\gamma}=A_{\al\be\gamma\delta}q_{\be}q_{\delta}
\end{equation}
already introduced in the wave equation  Eq.~\ref{wave_u} derived from the Mori-Zwanzig equations of motion. The summations over $\be, \delta$ in the expression for $\Lambda_{\al\gamma}$ indicate that its components are always symmetric combinations $(A_{\al\be\gamma\delta}+A_{\al\delta\gamma\beta})$. On the other hand, the tensor coefficients $B_{\al\be\gamma\delta}$ are measurable from linear response relations between stress and strain in simulations and experiments (see also Section~\ref{Ec_MD}). From the relations Eq.~\ref{A_B_C_rel}, it follows
\begin{equation}\label{A_B_sym_sum}
A_{\al\be\gamma\delta}+A_{\al\delta\gamma\beta}=B_{\al\be\gamma\delta}+B_{\al\delta\gamma\beta}
\end{equation}
For all practical measurements like speed of sound or response to mechanical deformation, $A_{\al\be\gamma\delta}$ or $B_{\al\be\gamma\delta}$ shows up as symmetric combinations of terms given in Eq.~\ref{A_B_sym_sum}.
Unlike $C_{\al\be\gamma\delta}$,  $A_{\al\be\gamma\delta}$ or $B_{\al\be\gamma\delta}$, do not have the full Voigt symmetry if they are measured for a pre-stressed reference lattice. However, if the reference system has purely isotropic pressure $\tau_{\al\be}=-P\delta_{\al\be}$, then $B_{\al\be\gamma\delta}$ becomes Voigt symmetric.

The number of independent components of the tensor $B_{\al\be\gamma\delta}$ depends on the symmetry of a lattice structure~\cite{WALLACE1970}. For the face centered cubic (FCC) crystal there are three independent components $B_{1111}, B_{1122}, B_{1212}$ which in Voigt notation are $B_{11}, B_{12}$ and $B_{44}$ respectively. This Voigt symmetric tensor ${\bf B}$ has a block diagonal form 
\begin{align}\label{mat_B}
&{\bf B}=
\begin{pmatrix}
B_{11} & B_{12} & B_{12}  \\
B_{12} & B_{11} & B_{12} & & \text{\huge0} \\
B_{12} & B_{12} & B_{11}\\
 && & B_{44}&0 & 0 \\
& \text{\huge0} & &0 & B_{44} & 0\\
 &&&0 & 0& B_{44} 
\end{pmatrix}
\end{align}
implying a decoupling between volume changing deformations associated with the top left diagonal block and pure shear with elastic constants related to $B_{44}$. The elastic moduli for bulk, bi-axial and shear deformations in FCC crystal can be derived from specific combinations of $B_{11}$, $B_{12}$ and $B_{44}$. In this paper, we will evaluate these quantities for a point-defect rich FCC crystal (i) analytically with input from classical density functional theory (section~\ref{Ec_cdft} and section~\ref{Ec_theo}) and (ii) from stress response to deformation experiments done in Molecular Dynamics (MD) simulations (section~\ref{Ec_MD} and section~\ref{sec_compare}).

\subsubsection{Elastic moduli from stress response to strain}\label{Ec_MD}
Theoretical predictions for elastic moduli from the free energy considerations of Sec.~\ref{th_fen} 
can be compared to results obtained from MD simulations using the following procedure.

When the equilibrium FCC crystal with Hamiltonian $\mathcal{H}$ is deformed by a small amount, this can be treated as a perturbation $\Delta \mathcal{H}$ modifying the Hamiltonian. Here, the stress and the strain are the pair of thermodynamic conjugate variables. Following arguments for linear response in presence of small perturbing fields, the stress response can be written in terms of the applied strain with $B_{\al\be\gamma\delta}$ acting as the proportionality constant. 
\begin{align}\label{Hookes}
\Delta \tau_{\alpha\beta}=\tau_{\al\be}(\epsilon_{\al\be})-\tau_{\al\be}(0)=&B_{\alpha\beta\gamma\delta}\epsilon_{\gamma\delta}
\end{align}
This essentially is the statement of Hooke's law. We examine three deformation protocols in our MD simulations (section~\ref{sec_compare}) where the simulation box is deformed by changing its shape. The elastic moduli obtained from the linear stress response in the MD simulation box subjected to deformation in the NVT ensemble are compared to the elastic moduli acquired  from the analytic evaluations (Section~\ref{Ec_theo}) of the second derivatives of the free energies with respect to coarse-grained elastic fields.
The general strain tensor with its nine components in 3D reduces to six independent components in case of a symmetric strain. For convenience of representation in the subsequent sections, we represent this as a 6 dimensional vector $(\epsilon_{11},\epsilon_{22},\epsilon_{33},\epsilon_{12},\epsilon_{13},\epsilon_{23})$.

\subsection{Elastic coefficients from the direct correlation function: input from classical density functional theory}\label{Ec_cdft}


In order to validate the theory proposed in Ref.~\cite{Walz2010} and summarised in Section~\ref{MZ_theo}, one needs to interpret the small wave vector $q\rightarrow 0$ and long time $t\rightarrow \infty$ limit for the equations (Eq.~\ref{eq_EOM}), giving static deformations. \gk{[in the following sentence it is not clear between what the context is established]} This provides the thermodynamics-based connection to the wave vector dependent correlations and the coarse-grained fields of elasticity theory derived from a microscopic starting point. 

According to classical density functional theory (DFT), for a reference crystalline solid in equilibrium with a microscopic density distribution $n(\br)$, the change in
free energy to second order in a density deviation $\delta\rho(\br)$ is given by\cite{Evans1979} 
\begin{align}\label{SF}
    \Delta F&=\dfrac{k_{B}T}{2}\int \int d^{3}r_{1}d^{3}r_{2}\left[\dfrac{\delta (\br_{12})}{n({\br_{1}})}-c(\br_{1},\br_{2})\right]\delta \rho(\br_{1})\delta \rho(\br_{2})
\end{align}
Here, $c(\br_{1},\br_{2})$ is the two-particle direct correlation function for the crystalline solid in equilibrium.
It is related to the inverse density correlation matrix $J_{\bg'\bg''}(\bq)$ (defined in Eq.~(\ref{eq_J})) through
 \begin{align}\label{jgg_crr}
{\bf J}_{\bg \bg^{'}}=&\frac{k_{B}T}{V}\int d^{3}r_{1}\int d^{3}r_{2} e^{i\bg\cdot\br_{1}}e^{-i\bg^{'}\cdot\br_{2}}e^{i\bq\cdot(\br_{1}-\br_{2})}\nonumber\\
&\left[\frac{\delta(\br_{1}-\br_2)}{n(\br_{1})}-c(\br_{1},\br_{2})\right]
\end{align}
which in turn leads to the dynamical matrix (see Eq.~(\ref{eq_Lambda}))
\begin{align}
    \Lambda(\bq)&=\sum_{\bg,\bg'}(g'+q)_{\alpha}n^{*}_{\bg'}J^{*}_{\bg'\bg}n_{\bg}(g+q)_{\beta}\nonumber\\
    &=\lambda_{\alpha \beta}({\bf q})-iq_{\alpha}\mu_{\beta}({\bf q})+iq_{\beta}\mu_{\alpha}^{*}({\bf q})+q_{\alpha}\nu({\bf q})q_{\beta} .
\end{align}
Here, the wavelength dependent generalised elastic coefficients $\lambda_{\al\be}(\bq), \mu_{\al}(\bq)$ and $\nu(\bq)$
and their small-$q$ limit are given by
\begin{subequations}\label{lam_mu_nu_1}
\begin{align}
&\lambda_{\al\be}(\bq)=-\sum_{\bg,\bg^{'}}ig^{'}_{\al}n^{*}_{\bg^{'}}J^{*}_{\bg^{'}\bg}(\bq)n_{\bg}ig_{\be}
=\lambda_{\alpha\beta\gamma\delta}q_{\gamma}q_{\delta}+\mathcal{O}(q^{3})\\
&\mu_{\al}(\bq)=\sum_{\bg,\bg^{'}}n^{*}_{\bg^{'}}J^{*}_{\bg^{'}\bg}(\bq)n_{\bg}ig_{\al}=i\mu_{\al\be}q_{\be}+\mathcal{O}(q^{2})\\
&\nu(\bq)=\sum_{\bg,\bg^{'}}n^{*}_{\bg^{'}}J^{*}_{\bg^{'}\bg}(\bq)n_{\bg}=\nu+\mathcal{O}(q^{2})
\end{align}
\end{subequations}
\vspace{2mm}
The symmetries of the generalised elastic constants $\nu$, $\mu_{\al\be}$, $\lambda_{\al\be\gamma\delta}$ 
and their relation to second order changes in free energy in terms of the coarse grained fields $\delta {\bf u}$ and $\delta c_{d}$ 
are discussed in detail in Ref.~\onlinecite{Walz2010} and Ref.~\onlinecite{Haring2015}.
In brief, the $\lambda$'s encode ``strain--strain'' free energy density changes $\propto \lambda_{\alpha\beta\gamma\delta} u_{\gamma\alpha} u_{\delta \beta}$, the $\nu$ describes ``density--density'' free energy density changes'' $\propto \nu (\delta n_{{\bf g}=0})^2 $
and the $\mu$'s describe a cross-term $\propto \mu_{\alpha\beta} u_{\beta\alpha} \delta n_{{\bf g}=0}$. 

Having access to the direct correlation function $c(\br_{1},\br_{2})$ as an input from classical DFT and exploiting the symmetries of this function for a crystalline solid, it is possible to derive general numerically tractable expressions for $\nu$, $\mu_{\al\be}$, $\lambda_{\al\be\gamma\delta}$. Thus for a model crystal of interest, it is possible to evaluate these quantities at desired density, temperature, point-defect concentration, etc. 

Given the lattice periodicity of a crystalline solid, the direct correlation function is invariant under global translations by any lattice vector ${\bf R}$, {\it i.e.} $c(\br_{1}+{\bf R},\br_{2}+{\bf R})=c(\br_{1},\br_{2})$. If the $c(\br_{1},\br_{2})$ is represented as a function of center of mass ${\bf s}_{12}=(\br_{1}+\br_{2})/2$ and relative distance $\br_{12}=(\br_{1}-\br_{2})$ then $c({\bf s}_{12}+{\bf R},\br_{12})=c({\bf s}_{12},\br_{12})$. The dependence on ${\bf s}_{12}$ can be expanded in reciprocal lattice vectors $\bg$. With $\bk$ being the Fourier conjugate of $\br_{12}$, the Fourier representation of $c$ is given by
\begin{equation}\label{crr_cg}
c({\bf s}_{12},\br_{12})=\sum_{\bg} e^{i\bg\cdot{\bf s}_{12}}c_{\bg}(\br_{12})=\sum_{\bg} \int \frac{d^{3}k}{(2\pi)^{3}}e^{i\bg\cdot{\bf s}_{12}}e^{i\bk\cdot\br_{12}}\tilde{c}_{\bg}(\bk) .
\end{equation}
Next we present expressions for the $\lambda_{\al\be}(\bf q)$, $\mu_{\al}(\bf q)$ and $\nu(\bf q)$ in terms of $\tilde{c}_{\bg}$ which we later use to evaluate these quantities for our model of interest (Eq.~\ref{fcc_phi_r}), the FCC cluster crystal. Substituting $J^{*}_{\bg'\bg}$ using Eq.~\ref{jgg_crr} in the expression for $\lambda_{\al\be}(\bf q)$ in Eq.~\ref{lam_mu_nu_1}a and utilising the expansion of the gradient of the average density distribution 
\begin{equation}
    \nabla_{\alpha}n(\br)=\sum_{\bg}ig_{\alpha}n_{\bg}e^{i\bg.\br}\label{gradft}
\end{equation}
in terms of the Bragg peak amplitudes $n_{\bg}$, one obtains
%
\begin{align}
\lambda_{\al \be}(\bq)
=&\frac{k_{B}T}{V}\int d^{3}r_{1}\int d^{3}r_{2}\nabla_{\al}n(\br_{1})\nabla_{\beta}n(\br_{2})e^{-i\bq.(\br_{1}-\br_{2})}\nonumber\\
&\left[\frac{\delta(\br_{1}-\br_2)}{n(\br_{1})}-c(\br_{1},\br_{2})\right]
\end{align}
 Upon using an equation derived by Lovett, Mou, Buff, and Wertheim (LMB~\cite{LMBW1}W~\cite{LMBW2}), 
\begin{equation}
    \frac{\nabla_{\al}(n(\br))}{n(\br)}=\int d^{3}r'c(\br,\br')\nabla_{\al}n(\br')\label{LMBW}
\end{equation}
$\lambda_{\al\be}$ can be rewritten -- using Eq.\ref{crr_cg} --  as
\begin{subequations}
\begin{align}
    \lambda_{\al\be}(\bq)=&\frac{k_{B}T}{V}\int d^{3}r_{1}\int d^{3}r_{2}\nabla_{\al}n(\br_{1})\nabla_{\beta}n(\br_{2})c(\br_{1},\br_{2})\nonumber\\
    &\left(1-e^{-i\bq\cdot(\br_{1}-\br_{2})}\right)\\
=&\frac{k_{B}T}{V}\sum_{\bg,\bg^{'},\bg^{''}}g^{'}_{\al}n^{*}_{\bg^{'}}n_{\bg}g_{\be} V\delta(\bg^{''}+\bg-\bg^{'})\nonumber\\
&\left[\tilde{c}_{\bg^{''}}(-\bg^{''}/2-\bg) -\tilde{c}_{\bg^{''}}(-\bg^{''}/2-\bg+\bq)\right]
\end{align}
\end{subequations}
subsequently -- and using Eq. (\ref{jgg_crr}) -- leading to 
\begin{subequations}\label{lam_mu_nu_2}
\begin{align}
\lambda_{\al\be}(\bq)=&k_{B}T\sum_{\bg,\bg^{'}}g^{'}_{\al}n^{*}_{\bg^{'}}n_{\bg}g_{\be} \nonumber\\
&\left[\tilde{c}_{\bg^{'}-\bg}\left(\frac{-\bg^{'}-\bg}{2}\right) -\tilde{c}_{\bg^{'}-\bg}\left(\frac{-\bg^{'}-\bg}{2}+\bq\right)\right]\label{lam_ab}\\
\mu_{\be}(\bq)
=&k_{B}T\sum_{\bg,\bg^{'}}n^{*}_{\bg^{'}}n_{\bg}ig_{\be}\nonumber\\ &\left[\tilde{c}_{\bg^{'}-\bg}\left(\frac{-\bg^{'}-\bg}{2}\right) -\tilde{c}_{\bg^{'}-\bg}\left(\frac{-\bg^{'}-\bg}{2}+\bq\right)\right]\\
\nu(\bq)
=&\frac{Nk_{B}T}{V}-k_{B}T\sum_{\bg,\bg^{'}}n^{*}_{\bg^{'}}n_{\bg}\tilde{c}_{\bg^{'}-\bg}\left(\frac{-\bg^{'}-\bg}{2}+\bq\right) .
\end{align}
\end{subequations}
Arguments similar to those in case of $\lambda_{\al\be}$, lead to the expressions for $\mu_{\be}$ and $\nu$ in Eq.\ref{lam_mu_nu_2}b and Eq.\ref{lam_mu_nu_2}c respectively.

In the limit  $q\rightarrow 0$, these generalised elastic coefficients result in the elastic constants
$B_{\al\be\gamma\delta}$.
\begin{equation}
   B_{\al\be\gamma\delta}=\lambda_{\al\gamma\be\delta}+\lambda_{\be\gamma\al\delta}-\lambda_{\al\be\gamma\delta}+\delta_{\al\be}\mu_{\gamma\delta}+\mu_{\al\be}\delta_{\gamma\delta}+\nu\delta_{\al\be}\delta_{\gamma\delta}
\end{equation}
From our calculations (see Section~\ref{Ec_theo}) we obtain all the coefficients of the matrix in Eq.~\ref{mat_B}. It has the symmetry corresponding to a FCC crystal with the expected block diagonal form and the three distinct elastic constants. These distinct elastic constants $B_{11}$, $B_{12}$ and $B_{44}$ (see the matrix, Eq.~(\ref{mat_B})) are obtained as the following combinations of the generalised elastic constants in the $q\rightarrow 0$ limit.

\begin{subequations}\label{EC}
\begin{align}
    &B_{11}=C_{11}-p=\lambda_{xxxx}+2\mu_{xx}+\nu\\
    &B_{12}=C_{12}+p=2\lambda_{xyxy}-\lambda_{xxyy}+\mu_{xx}+\mu_{yy}+\nu\\
    &B_{44}=C_{44}-p=\lambda_{xxyy} .
\end{align}
\end{subequations}

\section{Model and Simulation}\label{Simulation}
\subsection{The model: cluster crystals} 
\label{sec_model}
The system chosen for our study is a system with pairwise, ultrasoft interactions described by the generalised exponential model potential of index 4 (GEM-4)  
\begin{equation}\label{fcc_phi_r}
\phi(r)=e_{0} \exp[-(r/ r_{0})^4] .
\end{equation}
\gk{As shown in Ref.~\onlinecite{Mladek2006} and subsequent papers such a system forms -- despite the mutual repulsion of the particles -- stable clusters of overlapping particles. In well-explored regions of temperature and density the system forms (BCC or FCC) cluster crystals (see, e.g.,  phase diagram in Ref. \onlinecite{Mladek2006}): here these clusters (which are relatively homogeneous in their size) occupy regular BCC and FCC lattices, hence the name cluster crystals. These crystals have the remarkable property that -- upon an increase in density -- the lattice constant remains invariant while the occupation number increases.}
In this contribution we study the face centered cubic (FCC) phase.
To obtain crystal density profiles and the direct correlation function, we use a density functional for the free energy in mean--field approximation (highly accurate for this system, see Refs.~\onlinecite{Mladek2006,MladekJPCB2007}), given by
\begin{eqnarray}
    \beta {\cal F}[\rho] &=& \int d\br \;\rho(\br) [\ln (\rho(\br)\lambda^3)-1] + \nonumber \\
    & & \frac{1}{2} \int d\br \int d\br'
    \rho(\br) \rho(\br') (\beta \phi(|\br - \br'|)) \;. \label{eq_F}
\end{eqnarray}
Here, $\lambda$ is the thermal de Broglie length and $\beta=1/(k_B T)$ is the inverse temperature with $k_B$ denoting the Boltzmann constant. For the crystal phase, we use
an {\em ansatz} for the density profile
\begin{equation}\label{fcc_cc_nr}
\rho(\br)=n_{c}\left(\frac{\mathcal{A}}{\pi}\right)^{3/2}\sum_{\{{\bf R}\}}e^{-\mathcal{A}(\br-{\bf R})^{2}}=\sum_{\{{\bf R}\}}n_{cl}(\br-{\bf R})
\end{equation}
with
$\{{\bf R}\}$ being the FCC lattice vectors. $n_{cl}(\br-{\bf R})$ denotes the cluster density profile of any single cluster around a lattice site ${\bf R}$. 
The density profile is characterized by the average occupancy number $n_c$ and the width $\mathcal{A}$ associated with the density profile of a cluster peak. For a specified average density $n_0$ and temperature $T$, the free energy functional (Eq.~(\ref{eq_F})) 
is minimized with respect to $n_c$ and $\mathcal{A}$ to obtain the equilibrium profile $n(\br)$ with the equilibrium $\mathcal{A}$ and $n_{c}$ plugged in the expression for $\rho(\br)$ (Eq.~\ref{fcc_cc_nr}) for the crystal state.

The ultra-soft repulsion of the interaction potential allows for a range of ordered phases with multi-occupancies of the lattice sites as well as large fluctuations in the occupation numbers; \gk{thus the system can be considered as dominated by local defects}. While the fluctuations in the lattice occupancies can be interpreted as the motion of local defects~\cite{Haring2015}, these fluctuations preserve the long range order in the crystalline phase. Unlike conventional, singly occupied lattice structures with point-defect concentrations close to zero (such as the hard sphere system\cite{Pronk_Frenkel_2001}) this model has a relatively short time scale associated with the motion of local-defects\cite{coslovich2011hopping}.

\subsection{MD simulations of the cluster crystal}
\label{subsec:sim_cluster}
\ggr{The simulation methodology is similar as in previous contributions \cite{coslovich2011hopping,Mladek2006,shrivastav2020on,shrivastav2021yielding}.}
\gps{The \gpsch{GEM-4} potential, \gpsch{defined in Eq.~(\ref{fcc_phi_r})}, is truncated at a distance $r_{\rm c} = 2.2r_{0}$ and shifted to zero so that it vanishes from  $r_{\rm c}$ onwards. 
\ggr{We define a dimensionless temperature, density and time via $T^*=k_{\rm B}T/e_{0}$, $n_0^*=n_0 r_0^3$ and  $t^* = t\sqrt{e_{0}/m}/r_0$, where}
$m$ is the mass of particles.} 

\gk{We use the LAMMPS package \cite{plimpton1995fast} to perform non-equilibrium MD simulations in the NVT-ensemble. We consider the cluster system at three different densities, namely $n_{0}^* = 4.5, 6.5$ and $7.5$ for  \ggr{the  set of temperatures $T^* = \{0.4, 0.5, 0.6, 0.7, 0.8\}$}. From literature \cite{Mladek2006} it is known that the system assumes at these state points a stable FCC cluster phase, where each site of the FCC lattice is occupied by a cluster of overlapping particles. Data available in  literature \cite{Mladek2006,coslovich2011hopping,MladekJPCB2007} provide evidence that the average number of particles pertaining to a cluster, $n_{c}$, assumes for the considered state points a value $n_{c} \simeq 9, 13, 15$, corresponding to the densities ($n_{0}^* = 4.5, 6.5, 7.5$, respectively) and a lattice constant $l_{a} = 2 r_0$, $r_{0}$ being the unit of length in the system.
We consider ensembles of $2304, 3328$ and $3840$ particles \ggr{for the three densities, respectively}, corresponding thus to systems with $256$ clusters at each density. The temperature of the system is kept fixed via a dissipative particle dynamics (DPD) thermostat \cite{soddemann2003dissipative}. The equations-of-motion are integrated via the velocity-Verlet algorithm using an integration time step \ggr{$\Delta t^* = 0.005$} \cite{allen2017computer}.} 

\gps{For a chosen value of the density, the initial configurations of our simulations are ideal FCC cluster crystals where each lattice site is occupied by corresponding $n_{c}$, completely overlapping particles and assuming a lattice constant that is compatible with the chosen value of the density. Starting from these configurations, the system is equilibrated over $10^{6}$ MD steps at a temperature $\ggr{T^*} = 0.8$. This equilibrated system is further evolved, now at the desired temperature, over $5\times 10^{6}$ MD steps (where it has reached the diffusive regime) \cite{coslovich2011hopping,shrivastav2021yielding}, storing on a regular basis configurations in intervals of $10^{5}$ MD steps. These configurations then serve as independent initial configurations for subsequent simulations: from each of these state points, 50 independent simulation runs have been launched.}

\section{Results}\label{results}
\subsection{Elastic coefficients from the direct correlation function: application to the system of FCC cluster crystal}\label{Ec_theo}
In order to obtain the \ggr{elastic constants} (Eq.~\ref{EC}) for the FCC cluster crystal, we first need to evaluate the \ggr{$\bf q$--dependent generalised elastic coefficients} (Eq.~\ref{lam_mu_nu_2}). 
\ggr{From the mean--field free energy functional, Eq.~(\ref{eq_F}), we read off the direct correlation function} 
\begin{equation}\label{dcf_fcc_cc}
    c(\br_{1},\br_{2})=-\beta \phi(r), \ r=|\br_{2}-\br_{1}| .
\end{equation}
\ggr{which is simply the negative, dimensionless interaction potential (Eq.~\ref{fcc_phi_r}).} Approximating the \gk{direct correlation function of the crystal by} an isotropic, \ggr{liquid--like} function simplifies the expressions for the generalised elastic constants
\ggr{considerably since $\tilde c_{\bg}({\bf k})=0$ for $\bg \neq 0$ and $\tilde c_{\bg=0}({\bf k})=-\beta\tilde{\phi}({\bf k})=-\beta\tilde{\phi}_{\bg}(\bq)$ in  terms of the Fourier transform $\tilde\phi$ of the interaction potential $\phi$}. Utilising the explicit expressions for $n_{\bg}$ corresponding to the density distribution $n(\br)$ (Eq.~\ref{fcc_cc_nr} after equilibration) 
\begin{equation}\label{fcc_cc_ng}
    n_{\bg}=\frac{n_{c}}{V}e^{-\bg^{2}/4\mathcal{A}}
    \gb{\sum_{{\bf R}}e^{-i\bq^{'} \cdot {\bf R}}\delta(\bq^{'}-\bg)}
\end{equation}
and 
\ggr{the Fourier transformed direct correlation function}
in Eq.~\ref{lam_mu_nu_2} results in the following expressions
\begin{subequations}\label{lam_mu_nu_fcc}
\begin{align}
    \lambda_{\al \be}(\bq)/n^{*2}_{0}&=\sum_{\bg}g_{\al}e^{-\bg^{2}/2\mathcal{A}}g_{\be} \left[\tilde{\phi}_{\bg}\left(\bq\right)-\tilde{\phi}_{\bg}\left(0\right) \right]\nonumber\\
    &\overset{\bq\rightarrow0}\approx\lambda_{\al\be\gamma\delta} q_{\gamma}q_{\delta}/n^{*2}_{0}+\dots\\
    \mu_{\al}(\bq)/n^{*2}_{0}&=i\sum_{\bg}g_{\al}e^{-\bg^{2}/2\mathcal{A}} \left[\tilde{\phi}_{\bg}\left(\bq\right)-\tilde{\phi}_{\bg}\left(0\right) \right]\nonumber\\
    &\overset{\bq\rightarrow0}\approx i\mu_{\al\be}q_{\be}/n^{*2}_{0}+\dots\\
    \nu(\bq)/n^{*2}_{0}&=k_{B}T/n^{*}_{0}+\sum_{\bg}e^{-\bg^{2}/2\mathcal{A}}\tilde{\phi}_{\bg}\left(\bq\right)\nonumber\\
    &\overset{\bq\rightarrow0}\approx\nu/n^{*2}_{0}+\dots .
\end{align}
\end{subequations}
\ggr{Below we give results for the elastic constants in units of $n_0^*(n_0e_0)$.}
\ggr{The numerical evaluation requires the minimization of Eq.~(\ref{eq_F} (as described)}
to obtain the equilibrium lattice occupancy number $n_{c}$ and $\mathcal{A}$ \ggr{(note that $1/\sqrt{\mathcal{A}}$ is the peak width or particle localisation length).} 
This gives the average density profile (Eq.~\ref{fcc_cc_nr}) and hence $n_{\bg}$ (Eq.~\ref{fcc_cc_ng}) of the equilibrium FCC cluster crystal. Next, after taking the small $q$ limit (see Appendix ~\ref{app_lam_mu_nu_q0}) of the functions in Eq.~\ref{lam_mu_nu_fcc}, we perform the lattice sums over the reciprocal lattice vectors corresponding to the FCC lattice of desired density $n^{*}_{0}$.

\begin{figure}
    \includegraphics[width=7.5cm]{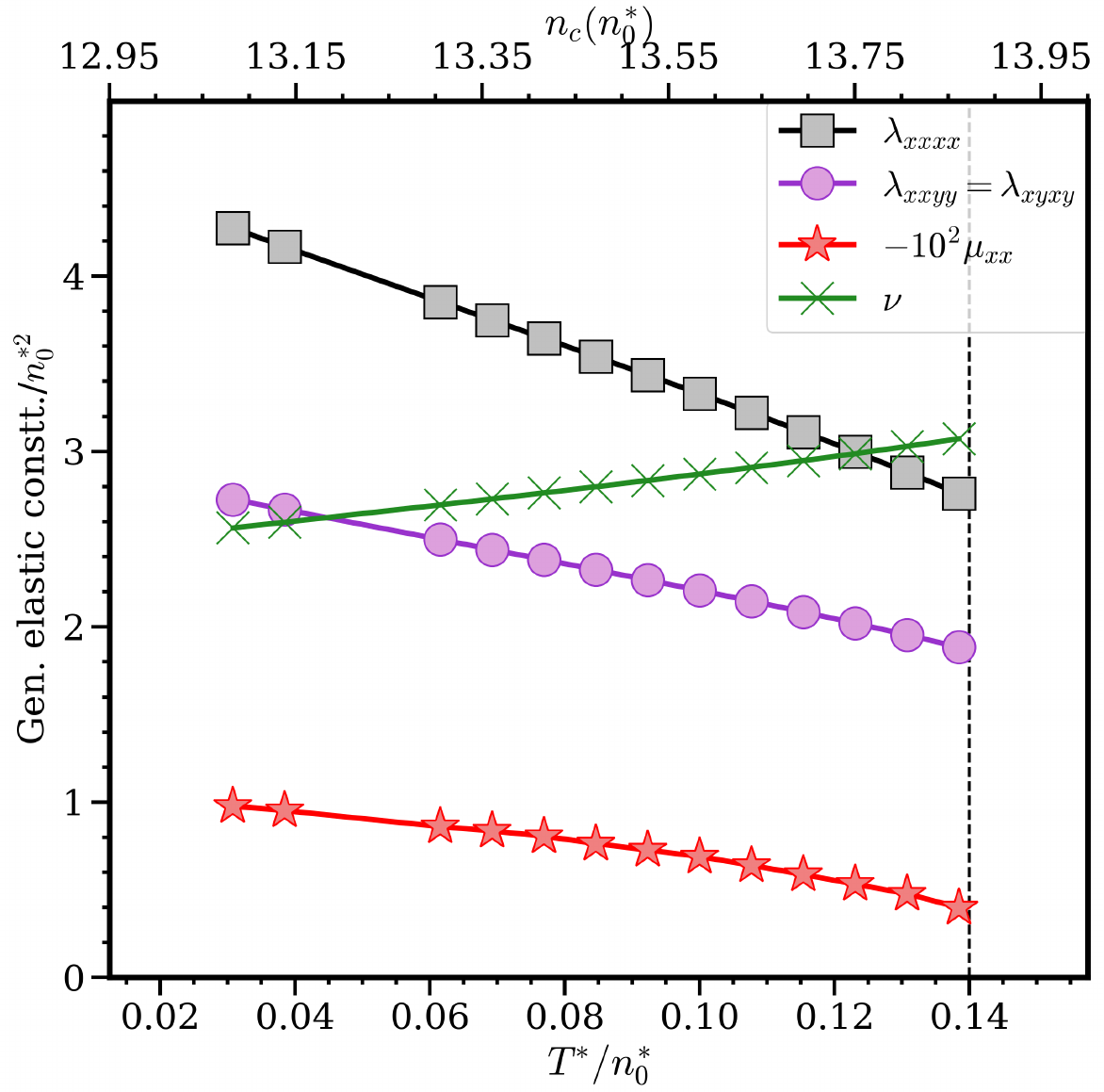}
    \caption{\label{GEc_fcc_cc} Generalised elastic constants in units of 
    \ggr{$n_0^* (n_0 e_0)$  for equilibrium FCC cluster crystals at $n_{0}^*=6.5$ for different temperatures, given as functions
    of $T^*/n_0^*$ ($x$--axis) and 
    $n_{c}$ (alternative $x$--axis). The vertical line at $T^*/n_0^*\approx 0.14$ separates the stable FCC cluster crystal phase and the body centered cubic (BCC) crystal phase.}}
\end{figure}
\begin{figure}
    \includegraphics[width=8.0cm]{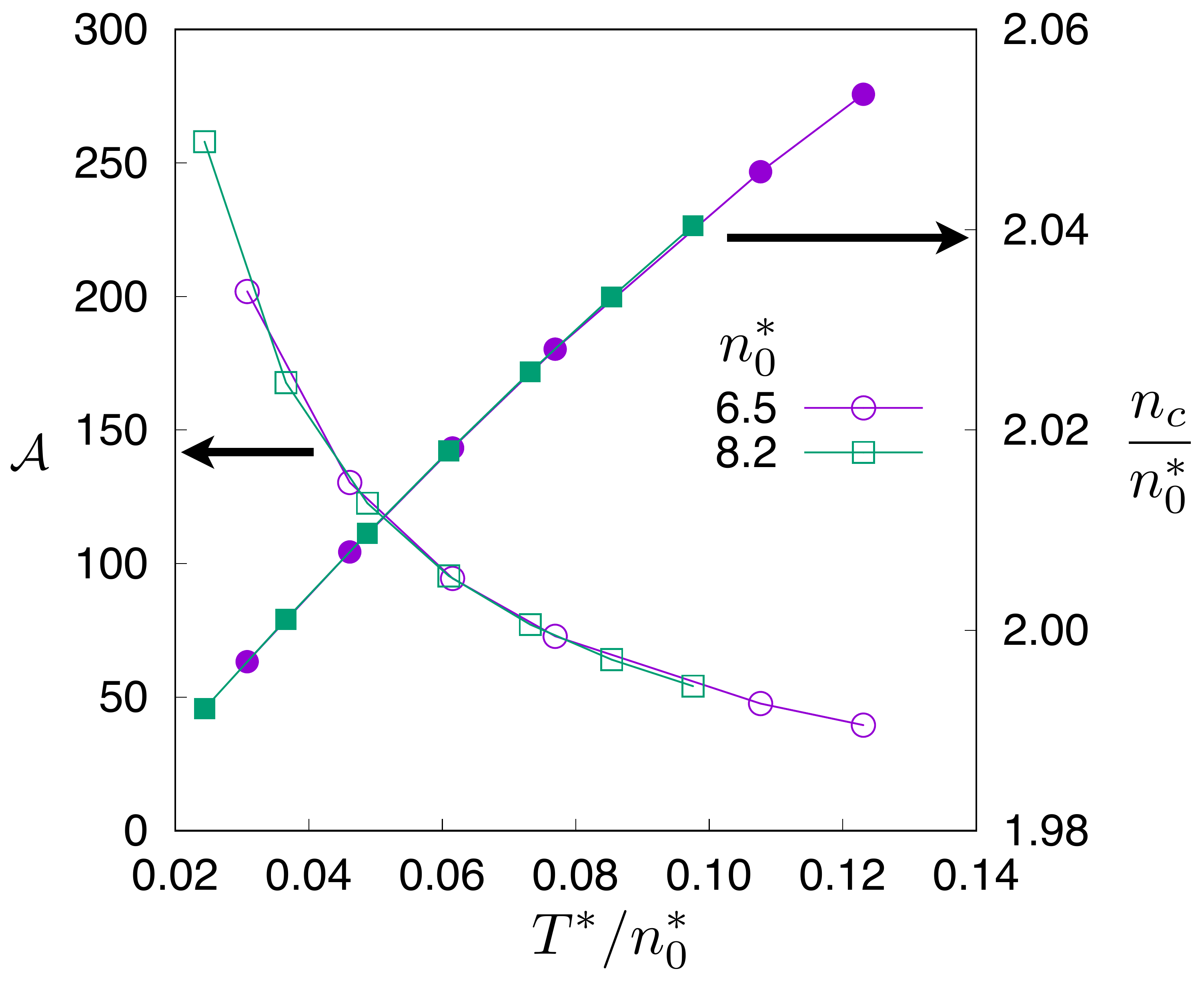}
    \caption{\label{A_Tbyn0}({\bf $y$-axis }) Plot of the cluster localisation parameter $\mathcal{A}$ 
    for equilibrium FCC cluster crystals for different $T^{*}/n^{*}_{0}$. The variances of the Gaussian distributions $n_{cl}(\br-{\bf R})$ of the average density profile (Eq.~\ref{fcc_cc_nr}) are inversely proportional to $\mathcal{A}$. Note that the particles are more localised close to the reference lattice sites ${\bf R}$ for higher $n^{*}_{0}$ and lower $T^{*}$. ({\bf alternative $y$-axis }) Plot of $n_{c}/n^{*}_{0}$ 
    as a function of $T^{*}/n^{*}_{0}$. It shows how $n_{c}/n^{*}_{0}$ remains approximately equal to 2 over the entire range of temperature for two different densities. As the lattice parameter of the FCC cluster crystal is given by $(4n_{c}/n^{*}_{0})^{1/3}$, this behaviour implies almost unchanged equilibrium lattice distances over the entire range of densities and temperatures corresponding to the stable FCC phase.} 
\end{figure}
\gk{In Ref~\onlinecite{Haring2015} it is argued}, that the generalised elastic constant $\lambda_{\al\be\gamma\delta}$ is sensitive to second order changes in free energy due to changes in symmetrised strain fields defined for the ordered solid. The quantity $\nu$, on the other hand, is associated with changes in the total density. $\mu_{\al\be}$ represents the coupling between average density and strain fields in the point-defect rich FCC lattice. Its consistent negative value is an indication of the conventional relation between the density fluctuations and the gradient of the displacement field (Eq.~\ref{ansatz1_real}). The much smaller magnitude of this term , compared to $\lambda_{\al\be\gamma\delta}$ and $\nu$ is also particularly interesting. Therefore, we examine this point in great detail in the Appendix ~\ref{app_lam_mu_nu_q0}. 
Our calculations (see Fig.~\ref{GEc_fcc_cc} and Eq.~\ref{lam_mu_nu_q0}) for this system reveal special symmetries like $\lambda_{\al\be\al\be}=\lambda_{\al\al\be\be}$ which is not a general property of the FCC lattice~\cite{lin2021}.

Careful observation of Eq.~\ref{lam_mu_nu_fcc} and consideration of Appendix ~\ref{app_lam_mu_nu_q0} shows how $\lambda_{\al\be\gamma\delta}$ and $\mu_{\al\be}$ vanishes for $\bg=0$ at $\bq\rightarrow 0$ leaving $\nu$ which is the only term to survive in a fluid with its ideal gas and an interaction contribution. The three distinct contributions in $\nu$ and how their combination is responsible for its increase with increase in temperature at constant density is discussed in Appendix ~\ref{app_lam_mu_nu_q0}.

Before trying to further interpret the dependence of the generalised elastic constants on \ggr{$T^*/n_0^*$},
one needs to clarify a special property of ordered structures with GEM-$4$ interaction potentials. The lattice parameter of the FCC cluster crystals is given by $(4n_{c}/n^{*}_{0})^{1/3}$ making the \gr{Cartesian components of a} reciprocal lattice vector $\bg$, at a given density, integral multiples of $2\pi(4n_{c}/n^{*}_{0})^{-1/3}$.
 The lattice parameter of the thermodynamically stable state is dictated by the position of the minimum in the Fourier transform of the interaction potential (see Fig.~\ref{exp_nu}b \ggr{below})~\cite{Likos2007}. Now, Fig.~\ref{A_Tbyn0} depicts how $n_{c}/n^{*}_{0}$ remains approximately constant at $2$ over the entire range of densities and temperatures thus making the $\bg$s almost independent of $n^{*}_{0}$. The localisation parameter $\mathcal{A}$, therefore, becomes the most influential quantity in dictating the mechanical response of the FCC cluster crystal. In this paper we verify this for one of these cluster crystal models but this expectation can be extended to the BCC cluster crystal structure as well.

The generalised elastic constants $\lambda_{\al\be\gamma\delta}$ and $-\mu_{\al\be}$ show an increase with decreasing temperature at a given density (Fig.~\ref{GEc_fcc_cc}). The following considerations will provide useful insights regarding this behaviour. An increase in 
\ggr{$T^*/n_0^*$} 
for the FCC cluster crystal also coincides with an increase in the equilibrium value of the average lattice occupancy $n_{c}$ (Fig.~\ref{GEc_fcc_cc}). Decreasing 
\ggr{$T^*/n_0^*$} 
on the other hand leads to an increase in $\mathcal{A}$ (Fig.~\ref{A_Tbyn0}) and hence more localisation of particles near lattice sites. A quantitative interpretation of this \gk{feature} can be derived by realising that the variance of the Gaussian distributions showing up in the expression for the average density profile (Eq.~\ref{fcc_cc_nr}) are proportional to $1/\mathcal{A}$. The predominant effect of the decrease in $\mathcal{A}$ with an increase in 
\ggr{$T^*/n_0^*$} 
(see Fig.~\ref{A_Tbyn0}) implies a greater \gk{propensity} of motion of particles across lattice sites. As the temperature $T^{*}$ is increased in the FCC cluster crystal while keeping density and hence the $\bg$s fixed, all the terms except $e^{-\bg^{2}/2\mathcal{A}}$ in Eq.~\ref{lam_mu_nu_fcc} are held constant. Therefore, a decrease in $\mathcal{A}$ results in a decrease in numerical contributions from the lattice sums in Eq.~\ref{lam_mu_nu_fcc}. This is also reflected in the temperature dependence of the overall elastic moduli of the FCC cluster crystals, related to  $\lambda_{\al\be\gamma\delta}$, $\mu_{\al\be}$ through Eq.~\ref{EC} and shown in Fig.~\ref{compare}.

Weaker localisation as a result of small $\mathcal{A}$, means an increased ease of hopping of particles or motion of local defects across lattice sites for the FCC cluster crystals. Thus a decrease in the elastic constants with a decrease in $\mathcal{A}$ provides a fundamental quantifiable basis to the intuitive expectation of a lower elastic moduli in a system with a higher probability of stress relaxation through diffusion of local defects.

The behavior of the generalized elastic constants for the cluster crystal is quite different from the ones in the hard sphere FCC crystal studied recently in Ref.~\onlinecite{lin2021}. In equilibrium, stable hard sphere crystals have vacancy concentrations $n_\text{vac}$ close to zero but one may study the behaviour of the generalized elastic constants as a function of defect density by artificially imposing a certain $n_\text{vac}$. It turns out that most of the generalized elastic constants diverge upon  $n_\text{vac} \to 0$, roughly as
$1/n_\text{vac}$. This behavior results from using a direct correlation function  $c(\br_{1},\br_{2})$ compatible with the full anisotropy of the lattice structure. If one uses a liquid--like direct correlation function $c(|\br_{1}-\br_{2}|)$, the dependence on $n_\text{vac}$ vanishes
almost completely. For hard spheres, the use of a liquid--like $c$ misses the singularities that are associated with local packing fractions at lattice sites going to 1; this is different for cluster crystals where the configuration at a lattice site is a collective state very well describable with a mean--field, liquid--like $c$. 

The generalised elastic constants evaluated in the small $q$ limit (see Fig.~\ref{GEc_fcc_cc}) and plugged into Eq.~\ref{EC} gives all the coefficients of elasticity which we compare to results from Molecular Dynamics simulations (see Fig.~\ref{compare}). These comparisons and their interpretations are detailed in Section~\ref{sec_compare}.

Having calculated the coefficients of elasticity, we use them to examine the speed of sound along different symmetry directions in the FCC cluster crystal. 

\begin{figure}[H]
	  \begin{center}
	  \includegraphics[width=7.5cm]{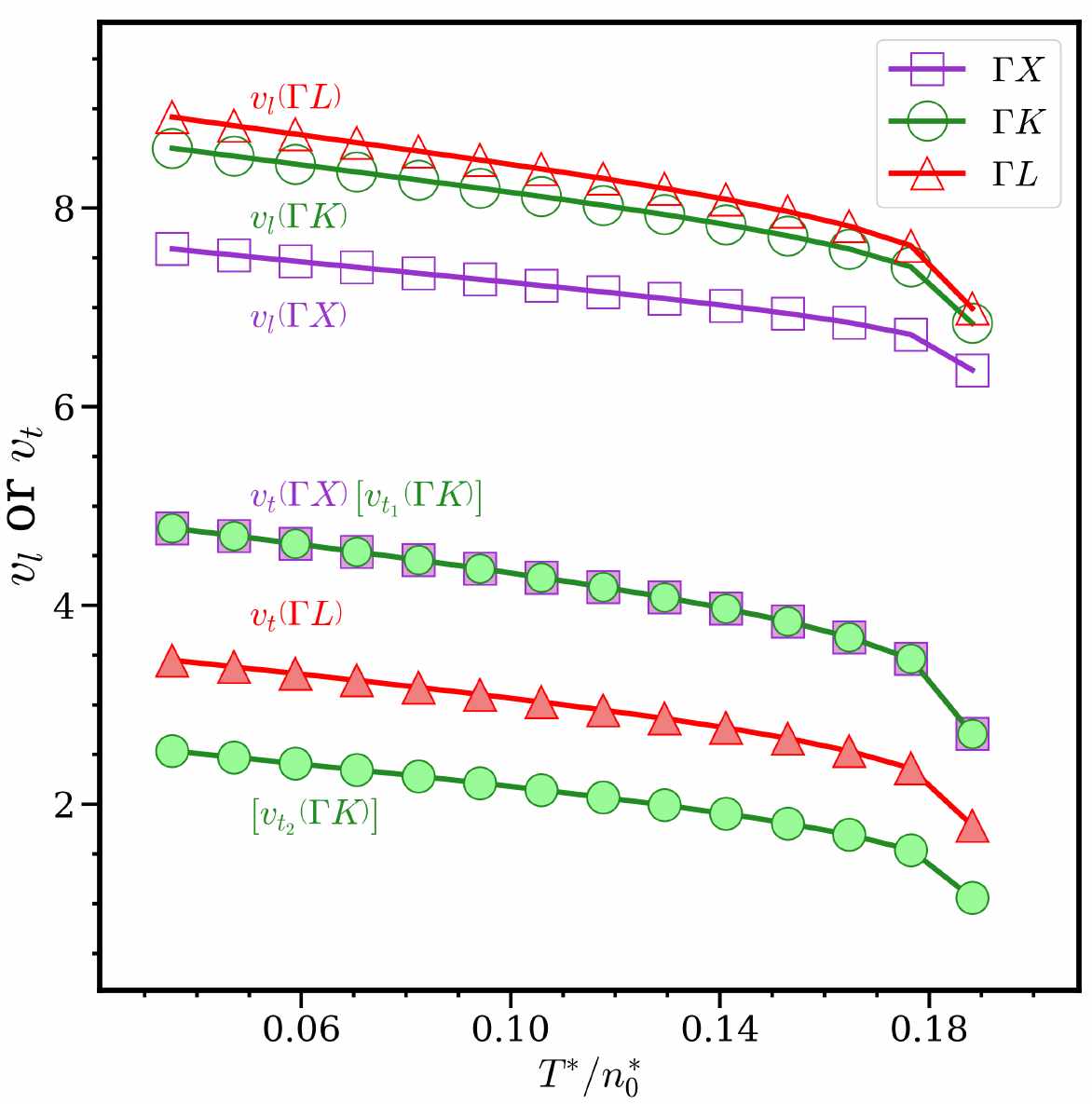} 
	  \caption{\label{speed_of_sound_FCC_cc}Transverse ($v_{t}$) and longitudinal ($v_{l}$) sound velocities along different symmetry directions obtained using classical DFT inputs for the FCC cluster crystal. $v_{l}$s are represented by open symbols while the $v_{t}$s are represented using filled symbols. Different colors are used to denote different directions of the ${\bq}$ vector. Purple squares represent the sound velocities along the symmetry direction $\Gamma X[100]$: $v_{l}=\sqrt{B_{11}/n^{*}_{0}}, \ v_{t}=\sqrt{B_{44}/n^{*}_{0}}$. Sound velocities along $\Gamma K[110]$ are shown with green circles: $v_{l}=\sqrt{(B_{11}+B_{12}+2B_{44})/2n^{*}_{0}}, \ v_{t1}=\sqrt{B_{44}/n^{*}_{0}}, \ v_{t2}=\sqrt{(B_{11}-B_{12})/2n^{*}_{0}}$. Finally the red triangles correspond to sound velocities along $\Gamma L[111]$: $v_{l}=\sqrt{(B_{11}+2B_{12}+4B_{44})/3n^{*}_{0}}, \ v_{t}=\sqrt{(B_{11}-B_{12}+B_{44})/3n^{*}_{0}}$}
	  \end{center}
\end{figure}

We observe that the velocities of sound increase monotonically with a decrease in temperature $T^{*}$ at a given density $n^{*}_{0}$. These results are a reiteration of the results presented in Fig.~\ref{GEc_fcc_cc} while bringing forth the obvious connection between the velocities of sound through an elastic medium and the elastic constants of that medium.

\subsection{Comparison : Theoretical Predictions and Simulations}\label{sec_compare}


\subsubsection{Shear modulus from volume preserving symmetric shear simulations}
\begin{figure*}
    \includegraphics[width=17cm]{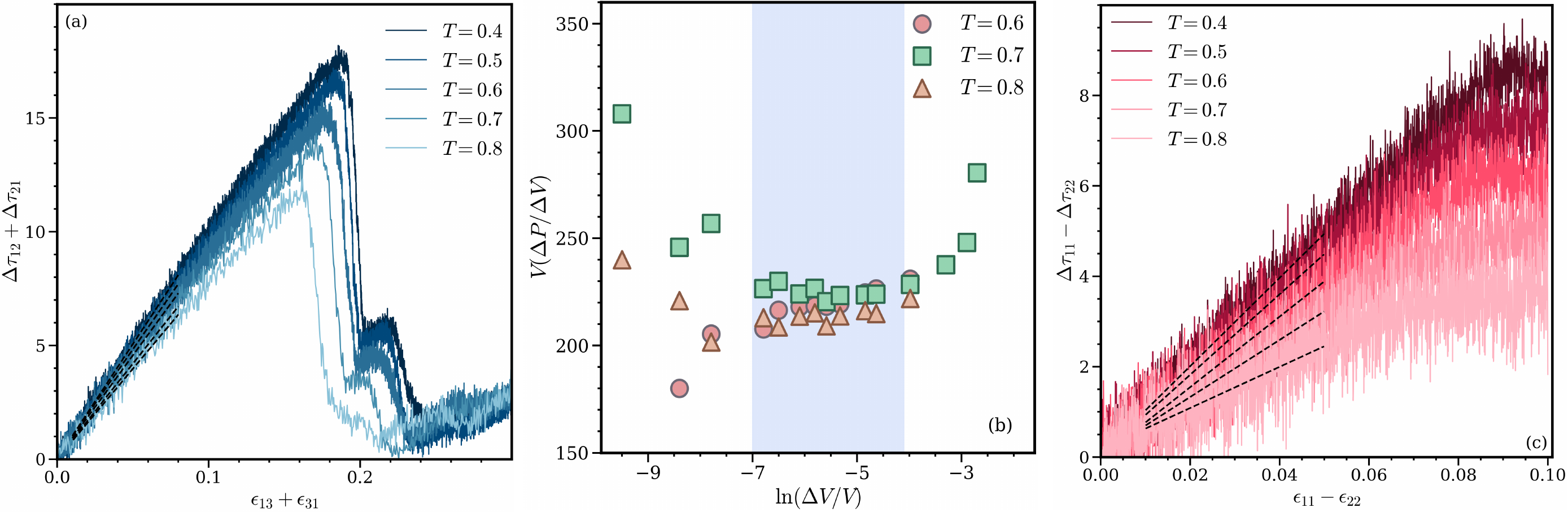}
    \caption{\label{stress_strain_LRT} (a) Plot of stress response to small deformations of the \gpsch{cluster crystal} system for five different temperatures as labelled. \gpsch{The black dashed lines represent the slopes of the initial linear regime, which gives the elastic modulus $B_{44}$.} (b) Variation of the bulk modulus (defined as \gr{$V(\Delta P^*/\Delta V$) with $P^*=Pr_0^3/e_0$} of the cluster crystals at density $n^{*}_{0}=6.5$ as a function of change in volume $\Delta V$ for three different temperatures as labelled. The elastic modulus $(B_{11}+2B_{12})/3$ is obtained by taking the average over the bulk modulus values in the blue shaded region.
    (c) Variation of the normal stress difference as a function of strain $\epsilon$ for the cluster crystal at density $n^{*}_{0} = 6.5$ at five different temperatures as labelled. \gpsch{The dashed black lines denote the slope of the initial linear regime, which defines the elastic modulus $2(B_{11}-B_{12})$.}} 
\end{figure*}

For a volume preserving symmetric shear $\epsilon_{13}=\epsilon_{31}=\epsilon$ of the $(x, z)$-plane with no change along the y-direction, the six-dimensional deformation vector is $(0,0,0,0,\epsilon,0)$. The shear stress response, conventionally represented as 
\begin{equation}
    (\Delta \tau_{12}+\Delta \tau_{21})=B_{1212}(\epsilon_{13}+\epsilon_{31})=B_{1212}(2\epsilon)
\end{equation}
is determined by the shear modulus $B_{1212}=B_{44}$.
\gps{\gk{In our simulations} we impose planar Couette flow on the bulk cluster crystal via Lees-Edwards boundary conditions \cite{lees1972computer}. The shear is applied in the $(x, z)$-plane along the $x$-direction; thus, the $z$- and $y$-directions are the gradient and vorticity directions, respectively, while $x$ is the shear-direction. The shear rate $\dot \gamma$ is considered to be equal to $\dot{\gamma} = 10^{-4}$. We note that for small deformations the slope of the linear regime does not depend on the shear rate.} This applied affine strain $\epsilon_{\al\be}$ results in a stress response in the system which is measured from the virial stress at a given temperature. The slope of these linear stress-strain curves allow us to calculate the elastic moduli. \gps{Fig.~\ref{stress_strain_LRT}(a) shows the stress-strain response of the cluster crystals at density $n^{*}_{0} = 6.5$ and temperatures $T^{*} = 0.4,0.5,0.6,0.7,0.8$. The black dashed lines represent the slope of the initial linear regime.}
\subsubsection{\gps{Bulk modulus from isotropic compression simulations}}\label{bm_MD}
In case of volume changing isotropic compression or expansion $\epsilon_{11}=\epsilon_{22}=\epsilon_{33}=\epsilon$ and $\epsilon_{12}=\epsilon_{13}=\epsilon_{23}=0$. $\epsilon$ is some small but finite deformation. The stress response or the change in pressure due to this deformation is reflected in the change in the trace of the stress tensor. Thus collecting all the terms via
\begin{subequations}\label{Hookes_2}
\begin{align}
\Delta \tau_{11}
=&B_{1111}\epsilon_{11}+B_{1122}\epsilon_{22}+B_{1133}\epsilon_{33}
\end{align}
\end{subequations}
the total stress response is 
\begin{equation}
    \dfrac{\left(\Delta \tau_{11}+\Delta \tau_{22}+\Delta \tau_{33}\right)}{3}=\dfrac{\left(B_{1111}+2B_{1122}\right)}{3}\epsilon
\end{equation}
leading to the bulk modulus $(B_{11}+2B_{12})/3$. Realising that $\tau_{\al\be}=-P\delta_{\al\be}$, the bulk modulus can be represented in terms of a change in pressure 
\begin{align}
    (B_{11}+2B_{12})/3=-V\left(\Delta P/\Delta V\right)\big|_{T^{*}} .
\end{align}
\gps{In our MD simulations, we consider the equilibrated cluster crystal system at different state points as mentioned above and slowly compress these samples. After each compression step the system is allowed to relax up to $10^{6}$ MD steps, a time which is larger than the typical equilibration time scales. \gpsch{The ratio of the resulting change in pressure $\Delta P$ and change in volume $\Delta V$ defines the bulk modulus, which is  plotted in Fig.~\ref{stress_strain_LRT}(b) as a function of change in volume $\Delta V$ for three different temperatures as labelled. The shaded blue region denotes the reliable linear regime of $\Delta V$ over which the bulk modulus remains nearly constant. At a given temperature, the bulk modulus is obtained by taking the average over the values of the bulk modulus in the blue region.}}
\subsubsection{\gps{Elastic moduli from bi-axial extension simulations}} \label{blam_MD}
In case of volume preserving, bi-axial deformation the six-dimensional deformation vector is $(\epsilon/2,-\epsilon/2,0,0,0,0)$. The deformation ($\epsilon_{11}=\epsilon/2$), is elongation along the $x$-axis of the lattice while ($\epsilon_{22}=-\epsilon/2$) defines a proportionate compression along $y$-axis thus preserving the volume of the simulation box up to linear order. The stress response produced by this deformation is given by 
\begin{equation}
    \dfrac{(\Delta\tau_{11}-\Delta\tau_{22})}{2}=(B_{1111}-B_{1122})(\epsilon_{11}-\epsilon_{22})=(B_{1111}-B_{1122})\epsilon .
\end{equation}
Therefore, the quantity $2(B_{1111}-B_{1122})=2(B_{11}-B_{12})$ can be \gk{extracted} from the slope of the stress-strain curves obtained in the MD simulations. 
\gps{To obtain these moduli from MD simulations, we perform a bi-axial deformation on the cluster crystal system at different densities and temperatures. The bi-axial deformation with a rate $10^{-4}$ is applied in the NVT ensemble along the $x$-direction preserving the volumes in the other two directions. The evolution of the normal stress difference $ \Delta\tau_{11} - \Delta\tau_{22}$ has been recorded as a function of the strain $\epsilon = \dot{\gamma}t$, see Fig.~\ref{stress_strain_LRT}(c). The slope of the stress-strain response in the linear regime yields the elastic moduli.}

\subsubsection{\gps{Comparison of the elastic moduli obtained from the theory and MD simulations}} \label{comp_Theo_MD}
Once we have determined  the different elastic moduli from the linear response relations between the stress-strain curves resulting from the deformation experiments performed via MD simulations, we compared these results with our theoretical predictions presented in Section~\ref{Ec_theo}.

\begin{figure}
    \includegraphics[width=8.0cm]{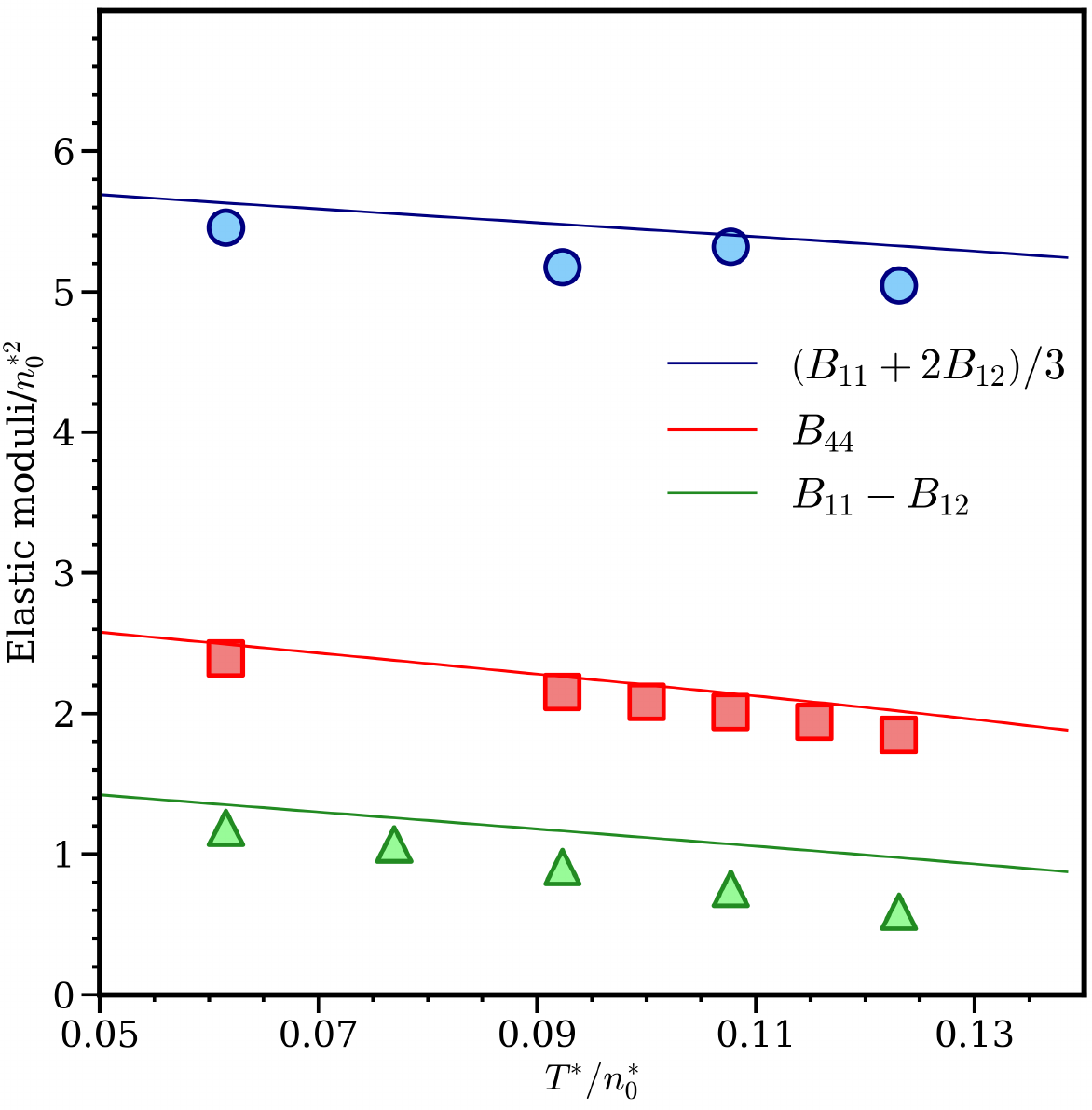}
    \caption{\label{compare} Comparison of theoretical predictions of elastic moduli to results obtained from linear stress response of FCC cluster crystal subjected to compression, shear and bi-axial deformation in Molecular dynamics simulations. The elastic constants are shown in units of 
    \gr{$n_0^*(n_0 e_0)$}
    as a function of the reduced thermodynamic parameter $T^{*}/n^{*}_{0}$. Note that the simulation results from different densities $n^{*}_{0}=6.5,...$ fall on the same linear curve when plotted in these reduced units.}
\end{figure}

Fig.~\ref{compare} shows the comparison of the bulk, bi-axial and shear moduli with the lines for theoretical predictions and bold symbols for simulation results. The error bars obtained from the variance of the simulation results provides an estimate for the accuracy of our simulation results. 
It is interesting to note the \gk{very good} agreement between the theoretical and simulation results despite the simplifying approximation of an isotropic direct correlation function (Eq.~\ref{dcf_fcc_cc}) for the FCC cluster crystal. The trend of a decrease in all the elastic moduli as $T^{*}/n^{*}_{0}$ increases can be explained from the similar trends observed in some of the generalised elastic constants (Section~\ref{Ec_theo}). Once again, the weaker particle localisation around lattice sites, quantified by smaller values of $\mathcal{A}$, accompanies the decrease in the elastic moduli. Since $\mathcal{A}$ proves to be an important factor in determining the elastic properties of the FCC cluster crystal, we try to associate the theoretically obtained localisation parameter (Section~\ref{sec_model}) with experimentally measurable quantities like the mean square displacement.
\subsubsection{\gps{The cluster localization parameter from MD simulations}}\label{LP_MD}
\begin{figure}
	  \begin{center}
	  \includegraphics[width=7.50cm]{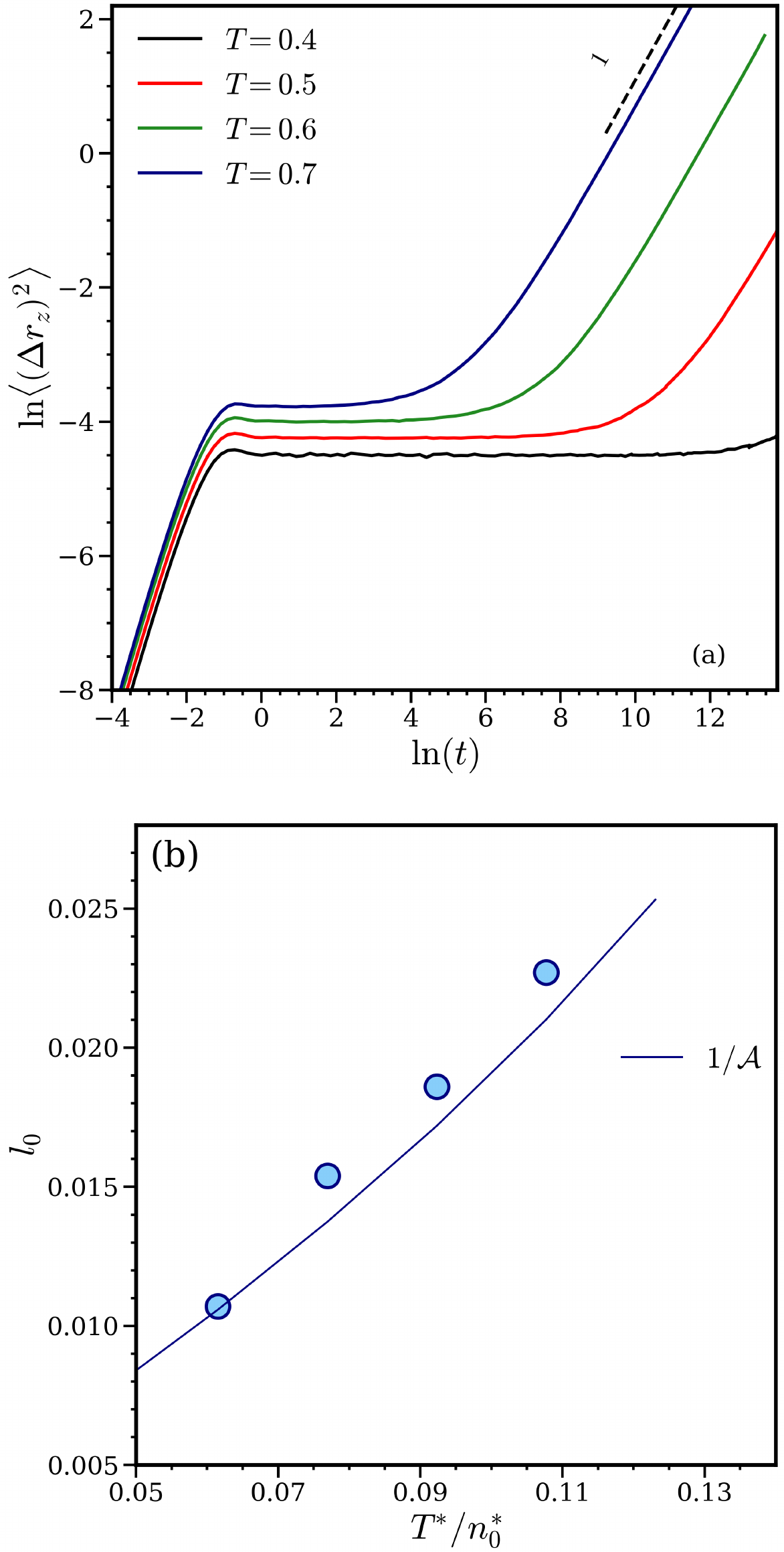} 
	  \caption{\label{msd_equib}\gps{(a) Equilibrium MSD \gr{(in units of $r_0^2$)} of particles of the cluster crystal system at $n^{*}_{0} = 6.5$ and temperatures $T^{*} =$ 0.4 (black), 0.5 (red), 0.6 (green), 0.7 (blue). The black dashed line represents a line with a slope equal to 1.} (b) Height of the plateau $l_{0}$ \gr{(in units of $r_0^2$)} in the MSD curves shown in Fig.~\ref{msd_equib} as a function of $T^{*}/n^{*}_{0}$. The solid line denotes the values of $1/\mathcal{A}$ obtained from Eq.~(\ref{fcc_cc_nr}). }
	  \end{center}
\end{figure}

\gps{In order to obtain the localization parameter $\mathcal{A}$ from MD simulations, we calculate the mean-square displacement (MSD) of particles, which is defined as,}
\begin{equation}
    \label{msd}
    \langle\Delta r^{2}(t)\rangle = \frac{1}{N}\sum_{i=1}^{N}\langle\left|\bm{r}_{i}(t) - \bm{R}^{0}_{i}\right|^2\rangle,
\end{equation}
\gps{where $\bm{r}_{i}(t)$ and ${\bm R}^{0}_{i}$ are the positions of the $i^{th}$ particle at time $t$ and $t = 0$, the time origin considered for the MSD calculations. The angular bracket corresponds to the averaging over the 
number of independently prepared samples. The MSD of particles for four different temperatures $T^{*} = 0.4, 0.5, 0.6, 0.7$ at a fixed  density $n^{*}_{0} = 6.5$ is shown in Fig.~\ref{msd_equib}(a). Note that all MSD curves have an initial ballistic regime with slope two, reach a plateau corresponding to the localization of particles in the clusters at high temperature a diffusive regime sets in where particles diffuse as they hop from cluster to cluster. To extract $\mathcal{A}$, we calculate the height of the plateau $l_{0}$ of the MSD curves for different temperatures. In Fig.~\ref{msd_equib}(b), we show the variation of the height of the plateau in the MSD curves shown in Fig.~\ref{msd_equib}(a) as a function of $T^{*}/n^{*}_{0}$ for the density $n^{*}_{0} = 6.5$. We compare our results with the values of $1/\mathcal{A}$ } obtained from Eq.~(\ref{fcc_cc_nr}), which is given  by the blue solid line. 
\section{Conclusions and Outlook}
Our theoretical approach successfully predicts mechanical properties of a lattice structure with a finite density of local-defects. The extent of this local disorder has been previously quantified using the variance in the lattice occupation number $\langle \Delta n^{2}_{c}\rangle$, see Ref.  \onlinecite{Mladek2006}. As explained in Ref.~\onlinecite{Haring2015}, this quantity can serve as a parameter to gauge the validity of the mean field free energy functional that predicts homogeneous lattice structures where particles hop between lattice sites. Following these arguments, we expect our theory to describe the cluster crystals accurately in the range of temperatures in the phase diagram where the lattice structures with integral occupation numbers are not the thermodynamic stable states~\cite{Wilding2013,Haring2015}.

The possibility of studying the mechanical properties of crystals with very large and vanishing concentrations of point-defects within a single framework opens up new directions for investigation. 
\gr{For a more systematic comparison between the FCC cluster crystal and the 
hard sphere FCC crystal, it would be desirable to have the fully anisotropic
direct correlation function for the former, to compare with the fully anisotropic
direct correlation function of the latter\cite{lin2021}.} 
But, the microscopic basis~\cite{Walz2010,Haring2015} (see Section~\ref{MZ_theo}) examined in this paper provides the conceptual understanding required to explore the distinct roles of the elastic fields and the couplings between them.

Application of this conceptual premise to study disordered binary crystals~\cite{Ras2020} provide another interesting avenue for exploring phenomenons like piezoelectricity associated with optical phonon modes~\cite{piezo}. Additional attention~\cite{flothesis,Mabi_1,Mabi_2} and further interpretation of the memory terms, which we have ignored here, can lead to further insights into the transport properties of the ordered solid. In this paper we \gr{have presented} the groundwork for applying such theoretical perspective to understand mechanical properties observed in atomistic simulations and experiments. Its importance lies in the clear association presented between second order changes in free energies due to elastic fields~\cite{Haring2015,Mladek2007} and elastic constants derived from deformation experiments in disorder dominated finite crystals at well defined temperatures (section~\ref{Ec_MD} and section~\ref{sec_compare}).

Theoretical and simulation studies attempting to explain mechanical response of crystalline solids have argued~\cite{Fleming_Cohen,Mladek2007} for the requirement of additional thermodynamic variables to account for the broken continuous symmetry. Our choice of the Bragg peak amplitudes, an experimentally measurable quantity, as the crystal order parameter to distinguish the displacement fields from the density fluctuations provides a general treatment to understand the microscopic basis for elastic response in non-ideal crystals. A natural extension of this would be to examine how one can include the interplay of local-defects and topological defects in dictating elastic and plastic response of crystals. 

Previous studies have employed spatial projection operator formalism~\cite{sas1}, within a statistical mechanics framework, to segregate length-scale dependent particle displacement modes responsible for (i)the elastic response or the affine deformations and (ii)particle rearrangements or non-affine deformations. A description of dislocation pre-cursors in an equilibrated ideal solid at finite temperature, emerges as a consequence~\cite{sas2,amHCN,popli} of this analysis. This approach provides the foundation for understanding the onset of plastic response~\cite{saspnas} and the origin of shear rate dependence of the yield-point~\cite{theoprl}, in an ideal crystal, using the language of discontinuous phase transition. Spatial projection of displacements onto affine and non-affine subspace also proves to be useful~\cite{sasmartens} in studying the transformation paths in martensitic transitions exhibiting Bane distortion and shuffle~\cite{KB}. The Mori-Zwanzig projection operator formalism, the microscopic frame for deriving the hydrodynamics of local-defect rich crystals in this paper, uses relaxation time-scales as the basis for identifying separate sets of variables. In the long wavelength limit, this, like the spatial projection operator formalism introduced in Ref.~\cite{sas1}, recovers the classical elasticity theory. Therefore, one of the future avenues for investigation will be an attempt to convergence these perspectives. That can lead to fundamental insights in the context of reversible and irreversible mechanical response in ideal as well as defect rich crystals subjected to deformation or undergoing structural transformations.
\begin{acknowledgments}
This work is supported by Deutsche Forschungsgemeinschaft through a D-A-CH grant FU 309/11-1 and OE 285/5-1, and the Austrian Funding Agency (FWF) under grant number I3846-N36. 
\end{acknowledgments}

\appendix

\section{The long-wavelength limit for the generalised elastic constants for the FCC cluster crystal}\label{app_lam_mu_nu_q0}


When the direct correlation function $c(\br_{1},\br_{2})$ for a crystalline structure is approximated with a liquid like isotropic $c(r)=-\beta \phi(r)$ (Eq.~\ref{fcc_phi_r}) as we have done for the FCC cluster crystal, Eq.~\ref{lam_mu_nu_2} simplifies to Eq.~\ref{lam_mu_nu_fcc}. In order to obtain the bulk elastic constants, the $q\rightarrow 0$ limit of the expressions in Eq.~\ref{lam_mu_nu_fcc} needs to be evaluated. The following Eq.~\ref{lam_mu_nu_q0} summarises the generalised elastic coefficients for the FCC cluster crystal in the small $q$ limit.  

\begin{widetext}
\begin{subequations}\label{lam_mu_nu_q0}
\begin{flalign}
&\lambda_{\al \be\gamma\delta}
=n^{*2}_{0}\sum_{\bg}g_{\al}e^{-\bg^{2}/2\mathcal{A}}g_{\be} \left(\delta_{\gamma\delta}D^{(1)}+g_{\gamma}g_{\delta}D^{(2)}\right)\\
&\mu_{\al\be}
= n^{*2}_{0}\sum_{\bg}g_{\al}g_{\be}e^{-\bg^{2}/2\mathcal{A}}D^{(1)}\\
&\nu
=n^{*}_{0}T^{*}+n^{*2}_{0}e^{-\bg^{2}/2\mathcal{A}}\tilde{\phi}_{\bg}\left(0\right)\bigg\rvert_{\bg=0}
+n^{*2}_{0}\sum_{\bg\neq 0}e^{-\bg^{2}/2\mathcal{A}}\tilde{\phi}_{\bg}\left(0\right)
\end{flalign}
\end{subequations}
The terms $D^{(1)}$ and $D^{(2)}$ are given by the following expressions
\begin{subequations}\label{D1_D2}
\begin{align}
&D^{(1)}=4\pi\int \left[\frac{r^{4}\cos(gr)}{(gr)^{2}}-\frac{r^{4}\sin(gr)}{(gr)^{3}}\right]\phi(r)dr\\
&D^{(2)}=4\pi \int  \left[\left\{-\frac{r^{6}\sin(gr)}{(gr)^{3}}-\frac{3r^{6}\cos(gr)}{(gr)^{4}}+\frac{3r^{6}\sin(gr)}{(gr)^{5}}\right\}\phi(r)dr\right]
\end{align}
\end{subequations}
and they are derived from the first and the second derivatives of $\tilde{\phi}_{\bg}$ with respect to components of $\bq$ in the limit $\bq\rightarrow 0$.
\begin{subequations}\label{eq_D1_D2}
\begin{align}
&\tilde{\phi}_{\bg}(q)=4\pi\int r^{2}\frac{\sin(|\bg+\bq|r)}{|\bg+\bq|r}\phi(r)dr\label{gem_4_q}\\
&\tilde{\phi}_{\bg}(q)-\tilde{\phi}_{\bg}(0)=\sum_{\al=1}^{3}\frac{\partial \tilde{\phi}_{\bg}(q)}{\partial q_{\al}}\bigg\rvert_{\bq=0}q_{\al}+\frac{1}{2}\sum_{\al=1}^{3}\sum_{\be=1}^{3}\frac{\partial^{2} \tilde{\phi}_{\bg}(q)}{\partial q_{\al}\partial q_{\be}}\bigg\rvert_{\bq=0}q_{\al}q_{\be}+\dots \label{taylor_exp}\\
&\frac{\partial\tilde{\phi}_{\bg}(q)}{\partial q_{\al}}=\frac{\partial\tilde{\phi}(q^{\prime})}{\partial q_{\al}}=\frac{d\tilde{\phi}(q^{\prime})}{dq^{\prime}}\frac{\partial q^{\prime}}{\partial q_{\al}}=\frac{d\tilde{\phi}(q^{\prime})}{dq^{\prime}}\frac{q_{\al}^{\prime}}{q^{\prime}}, \ (q^{\prime}=|\bg +\bq|)\\
&\frac{\partial\tilde{\phi}(q^{\prime})}{\partial q_{\al}}\big|_{\bq= 0}=4\pi g_{\al}\int \left[\frac{r^{4}\cos(gr)}{(gr)^{2}}-\frac{r^{4}\sin(gr)}{(gr)^{3}}\right]\phi(r)dr=g_{\al}D^{(1)}\\
&\frac{\partial^{2}\tilde{\phi}(q^{\prime})}{\partial q_{\al}\partial q_{\be}}\big|_{\bq= 0}=4\pi \int  \left[\delta_{\al\be}\left\{\frac{r^{4}\cos(gr)}{(gr)^{2}}-\frac{r^{4}\sin(gr)}{(gr)^{3}}\right\}+g_{\al}g_{\be}\left\{-\frac{r^{6}\sin(gr)}{(gr)^{3}}-\frac{3r^{6}\cos(gr)}{(gr)^{4}}+\frac{3r^{6}\sin(gr)}{(gr)^{5}}\right\}\right]\phi(r)dr\\
&=\delta_{\al\be}D^{(1)}+4\pi \int  \left[g_{\al}g_{\be}\left\{-\frac{r^{6}\sin(gr)}{(gr)^{3}}-\frac{3r^{6}\cos(gr)}{(gr)^{4}}+\frac{3r^{6}\sin(gr)}{(gr)^{5}}\right\}\right]\phi(r)dr=\delta_{\al\be}D^{(1)}+g_{\al}g_{\be}D^{(2)}
\end{align}
\label{gem_4_Taylor}
\end{subequations}

\begin{figure*}[h]
    \includegraphics[width=15.0cm]{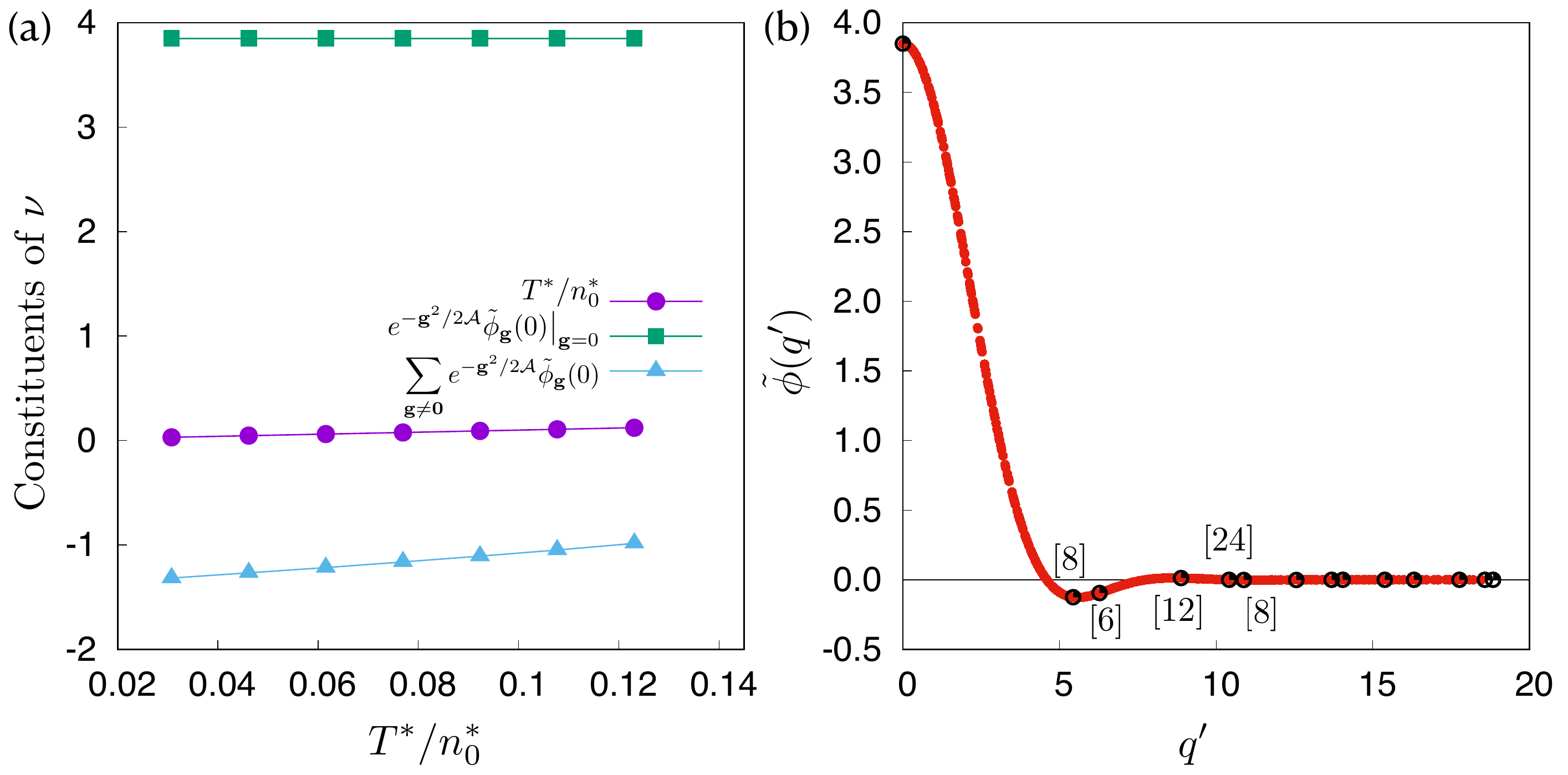}
    \caption{\label{exp_nu}(a) The three terms contributing to the generalised elastic constant $\nu$ for the FCC cluster crystal plotted in units of $n^{2}_{0}e_{0}r^{3}_{0}$ as a function of the reduced thermodynamic parameter $T^{*}/n^{*}_{0}$. The ideal gas term $T^{*}/n^{*}_{0}$ represented by purple circles with line, increases with $T^{*}/n^{*}_{0}$. The second term $e^{-\bg^{2}/2\mathcal{A}}\tilde{\phi}_{\bg}(0)\big|_{\bg=0}$ (green squares) provide a $T^{*}/n^{*}_{0}$ independent contribution arising from interaction terms at $\bg=0$. The third contribution $\sum_{\bg\neq 0}e^{-\bg^{2}/2\mathcal{A}}\tilde{\phi}_{\bg}(0)$ is a sum over all lattice vectors $\bg\neq 0$. This has a negative value with a magnitude decreasing with $k_{B}T/e_{0}n^{*}_{0}r^{3}_{0}$. The second and the third contributions can be understood by looking at the plot of $\tilde{\phi}(q')=\tilde{\phi}_{\bg}(q)$ in the figure on the right (b)Plot of $\tilde{\phi}(q')$ as a function of $q'$. The circles indicate the position of the neighbour shells in the reciprocal lattice for the FCC lattice. $\tilde{\phi}(q')$ is the Fourier transform of $\phi(r)$ (Eq.\ref{fcc_phi_r}). 
    Note the large positive magnitude of $\tilde{\phi}(0)$ showing up in the second term (Eq.~\ref{lam_mu_nu_q0}c) in $\nu$. The negative value of the lattice sum in the third term (Eq.~\ref{lam_mu_nu_q0}c) is justified by the sum over $\bg\neq 0$ indicated by the larger negative magnitudes of $\tilde{\phi}(q')$ marked by circles in this plot.}
\end{figure*}

\begin{figure*}[h]
    \includegraphics[width=15.0cm]{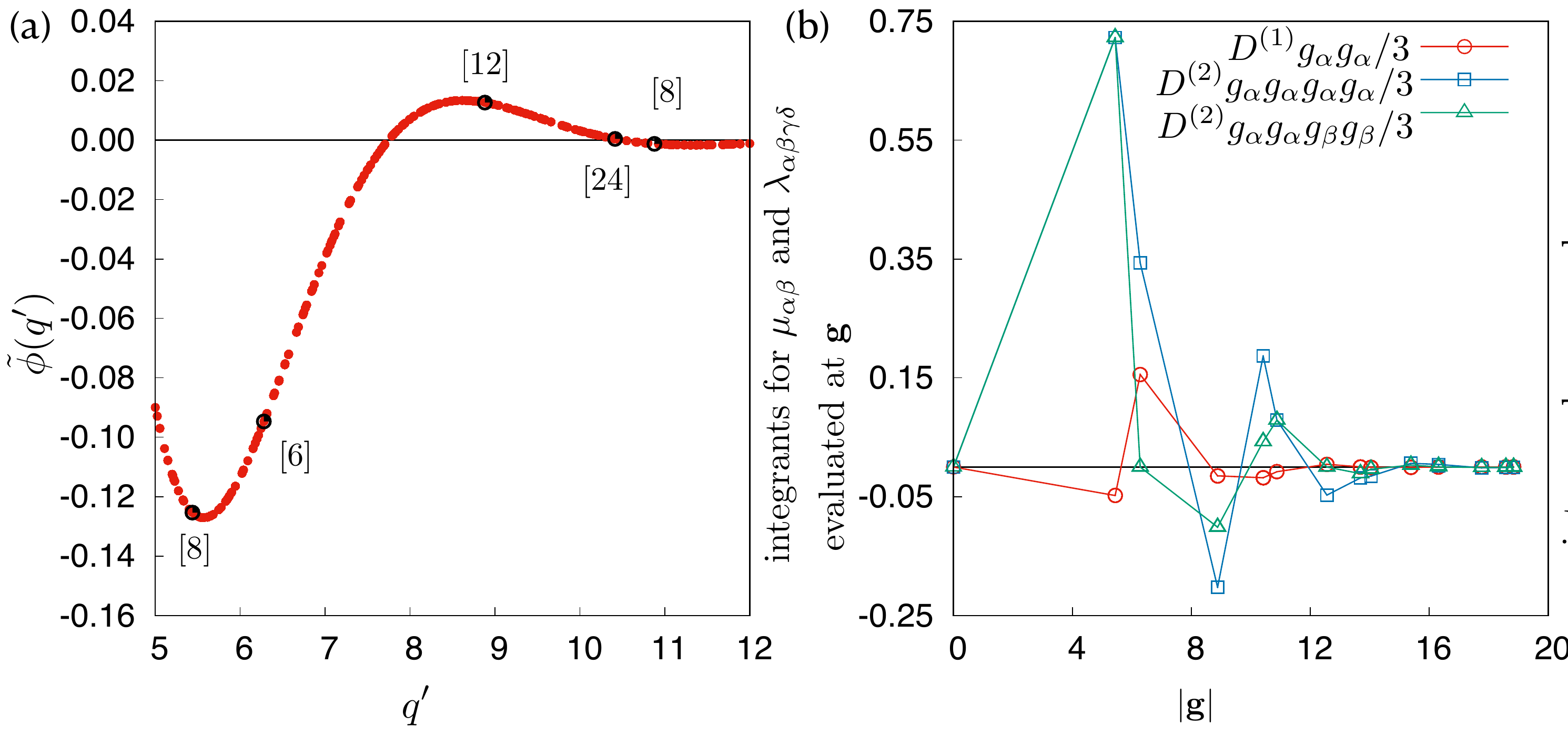}
    \caption{\label{exp_mu} (a)A zoomed in plot of $\tilde{\phi}(q')$ as a function of $q'$. The circles indicate the position of the neighbour shells in the reciprocal lattice for the FCC lattice. The reciprocal lattice shells with the largest contributions to the lattice sums in the expressions for $\lambda_{\al\be\gamma\delta}, \mu_{\al\be}$ and $\nu$ (Eq.~\ref{lam_mu_nu_q0}) are specially denoted with the exact number of lattice sites in those shells.  
    Note how the first neighbour shell is slightly shifted to the left of the position of the minimum in the curve.
     (b)Plot of the averaged $D^{(1)}(g^{2}_{x}+g^{2}_{y}+g^{2}_{z})/3$ as a function of the positions of the reciprocal lattice vectors $|{\bf g}|$ indicated by the red open circle symbols. This shows the contributions of each reciprocal lattice shell in the lattice sum in the expression for $\mu_{\al\be}$ in Eq.~\ref{lam_mu_nu_q0}b. The averaging over the co-ordinate directions are possible because of the symmetry $\mu_{xx}=\mu_{yy}=\mu_{zz}$. Blue open squares denote $D^{(2)}(g^{4}_{x}+g^{4}_{y}+g^{4}_{z})/3$ while green triangles represent $D^{(2)}(2g^{2}_{x}g^{2}_{y}+2g^{2}_{y}g^{2}_{z}+2g^{2}_{z}g^{2}_{x})/6$ as a function of  $|{\bf g}|$. These are the quantities that contribute to the lattice sums in the expression for $\lambda_{\al\be\gamma\delta}$ in Eq.~\ref{lam_mu_nu_q0}a. Here the averages are made possible by the symmetries $\lambda_{xxxx}=\lambda_{yyyy}=\lambda_{zzzz}$, $\lambda_{xxyy}=\lambda_{yyzz}=\lambda_{zzxx}$ and $\lambda_{xxyy}=\lambda_{xyxy}$.} 
\end{figure*}
\end{widetext}

\subsection{Understanding the generalised elastic constant $\nu$}
Now, if only $\bg=0$ is considered, Eq.\ref{lam_mu_nu_q0} shows that $\lambda_{\al\be\gamma\delta}$ and $\mu_{\al\be}$ vanishes. The term $\nu$ survives with the ideal gas contribution (the first term in Eq.\ref{lam_mu_nu_q0}c) and an interaction contribution from the second term in Eq.\ref{lam_mu_nu_q0}c. The second term with
\begin{align}
    &\tilde{\phi}_{\bg}(q)\big|_{\bg+\bq\rightarrow 0}=4\pi\int r^{2}\phi(r)dr\label{gem_4_q_0}
\end{align}
is a density and temperature independent constant in this limit. This is consistent with the elastic property of a homogeneous fluid medium where the bulk compressibility $(1/\nu)$ is obtained from the density fluctuations with no further contributions arising from the displacement fields defined in an ordered medium $(\lambda_{\al\be\gamma\delta})$ or its coupling to the density field $(\mu_{\al\be})$. 

Unlike, $\lambda_{\al\be\gamma\delta}$ and $-\mu_{\al\be}$, $\nu$ increases with an increase in $T^{*}/n^{*}_{0}$. Let us consider the individual terms contributing to $\nu$, for an ordered solid, as shown in Fig.~\ref{exp_nu}a. The ideal gas term $T^{*}/n^{*}_{0}$ increases with $T^{*}/n^{*}_{0}$ having a slope of one, as expected. The second term $e^{-\bg^{2}/2\mathcal{A}}\tilde{\phi}_{\bg}(0)\big|_{\bg=0}$ provides a $T^{*}/n^{*}_{0}$ independent contribution arising from interaction terms at $\bg=0$ as explained in the previous paragraph. The third contribution $\sum_{\bg\neq 0}e^{-\bg^{2}/2\mathcal{A}}\tilde{\phi}_{\bg}(0)$ is a sum over all lattice vectors $\bg\neq 0$. This has a negative value with a magnitude decreasing with increasing $T^{*}/n^{*}_{0}$. The second and the third contributions really comes from the values of $\tilde{\phi}(q')=\tilde{\phi}(|\bg+\bq|)$,  the Fourier transform of $\phi(r)$ (Eq.\ref{fcc_phi_r}), evaluated at the reciprocal lattice vectors $\bg=0$ and all the $\bg\neq 0$ respectively. The positions and the magnitudes of these terms are indicated on the plot of $\tilde{\phi}(q')$ in Fig.~\ref{exp_nu}b.

\subsection{Understanding the generalised elastic constant $\mu_{\al\be}$}
For the generalised elastic constant $\mu_{\al\be}$, the symmetry of the FCC lattice dictates (see Eq.~\ref{lam_mu_nu_q0}b) that all the cross-terms like $\mu_{xy}$ must be zero and $\mu_{xx}=\mu_{yy}=\mu_{zz}$. Here we examine why $\mu_{\al\al}$ has a much smaller magnitude compared to the other generalised elastic constants $\lambda_{\al\be\gamma\delta}$ and $\nu$. The expression in Eq.~\ref{lam_mu_nu_q0}b shows that the lattice sum in $\mu_{\al\al}$ relies on the magnitude of the function $D^{(1)}$ which is the first derivative of $\tilde{\phi}(q')=\tilde{\phi}(|\bg+\bq|)$ with respect to $q$ (see Eq.~\ref{eq_D1_D2}d) at $q\rightarrow 0$. A closer look at the zoomed in plot of $\tilde{\phi}(q')$ in Fig.~\ref{exp_mu}a, with the first $\bg\neq 0$ reciprocal lattice position slightly shifted from the minimum in $\tilde{\phi}(q')$, clearly shows why the $D^{(1)}$ evaluated here has a negative value. Following this argument, one immediately sees the justification of the magnitudes and signs of $D^{(1)}(g^{2}_{x}+g^{2}_{y}+g^{2}_{z})$ plotted at the positions of the reciprocal lattice vectors in Fig.~\ref{exp_mu}b. In this figure the contributing terms to $\mu_{\al\be}$ (Eq.~\ref{lam_mu_nu_q0}b) are compared to those in $\lambda_{\al\be\gamma\delta}$ (Eq.~\ref{lam_mu_nu_q0}a). The function $D^{(2)}$ (Eq.~\ref{D1_D2}b), the leading term in the lattice sum in $\lambda_{\al\be\gamma\delta}$, is related to the second derivative of $\tilde{\phi}(q')$ (see Eq.~\ref{eq_D1_D2}f). The curvature of $\tilde{\phi}(q')$ close to its minimum and arguments similar to those given in case of $\mu_{\al\be}$ can justify the signs and magnitudes of various $D^{(2)}$ dependent terms plotted in Fig.~\ref{exp_mu}b. Finally, a cumulative sum over the appropriately weighted data required to evaluate $\lambda_{\al\be\gamma\delta}$, $\mu_{\al\be}$ results in cancellations of terms in case of $\mu_{\al\al}$ leading to a much smaller value of $\mu_{\al\al}$ compared to $\lambda_{\al\be\gamma\delta}$. It also shows how these lattice sums saturate at around or before the thirteenth reciprocal lattice shell.



\bibliography{aipsamp}

\end{document}